\documentclass{article}
\usepackage{amsmath}
\usepackage{authblk}
\usepackage{lineno}
\usepackage[usenames,dvipsnames,svgnames,table]{xcolor}
\usepackage{graphicx}
\usepackage{subcaption}
\usepackage{lineno}
\usepackage{mathrsfs}
\usepackage{longtable}
\usepackage{multirow}
\usepackage{hyperref}
\usepackage[a4paper,margin=1in]{geometry}
\usepackage{cite}
\usepackage{rotfloat}
\newcommand{\tabincell}[2]{\begin{tabular}{@{}#1@{}}#2\end{tabular}} 
\begin{document}
%\begin{CJK*}{UTF8}{song}
%\title{Determination of responses of the PandaX-II liquid xenon detector to low energy electron and nuclear recoils}
\title{Determination of responses of liquid xenon to low energy electron and nuclear recoils using the PandaX-II detector}

\author[1,4]{Binbin Yan \thanks{corresponding author, yanbinbin@sjtu.edu.cn}}
\author[1]{Abdusalam Abdukerim}
\author[1]{Wei Chen}
\author[1,4]{Xun Chen \thanks{corresponding author, chenxun@sjtu.edu.cn}}
\author[5]{Yunhua Chen}
\author[6]{Chen Cheng}
\author[7]{Xiangyi Cui}
\author[8]{Yingjie Fan}
\author[9]{Deqing Fang}
\author[9]{Changbo Fu}
\author[10]{Mengting Fu}
\author[11,12]{Lisheng Geng}
\author[1]{Karl Giboni}
\author[1]{Linhui Gu}
\author[5]{Xuyuan Guo}
\author[1]{Ke Han}
\author[1]{Changda He}
\author[1]{Di Huang}
\author[15]{Peiyao Huang}
\author[5]{Yan Huang}
\author[13]{Yanlin Huang}
\author[1]{Zhou Huang}
\author[14]{Xiangdong Ji}
\author[15]{Yonglin Ju}
\author[7]{Shuaijie Li}
\author[1,7,4]{Jianglai Liu\thanks{spokesperson, jianglai.liu@sjtu.edu.cn}}
\author[15]{Zhuoqun Lei}
\author[1]{Wenbo Ma}
\author[9,2]{Yugang Ma}
\author[10]{Yajun Mao}
\author[1,4]{Yue Meng}
\author[1]{Kaixiang Ni}
\author[5]{Jinhua Ning}
\author[1]{Xuyang Ning}
\author[16]{Xiangxiang Ren}
\author[5]{Changsong Shang}
\author[1]{Lin Si}
\author[11]{Guofang Shen}
\author[14]{Andi Tan}
\author[16]{Anqing Wang}
\author[2,17]{Hongwei Wang}
\author[16]{Meng Wang}
\author[3,2]{Qiuhong Wang}
\author[10]{Siguang Wang}
\author[6]{Wei Wang}
\author[15]{Xiuli Wang}
\author[1,4]{Zhou Wang}
\author[6]{Mengmeng Wu}
\author[5]{Shiyong Wu}
\author[1]{Weihao Wu}
\author[1]{Jingkai Xia}
\author[14,18]{Mengjiao Xiao}
\author[7]{Pengwei Xie}
\author[15]{Rui Yan}
\author[1]{Jijun Yang}
\author[1]{Yong Yang}
\author[8]{Chunxu Yu}
\author[16]{Jumin Yuan}
\author[1]{Ying Yuan}
\author[5]{Jianfeng Yue}
\author[1]{Xinning Zeng}
\author[14]{Dan Zhang}
\author[1,4]{Tao Zhang}
\author[1,4]{Li Zhao}
\author[13]{Qibin Zheng}
\author[5]{Jifang Zhou}
\author[1]{Ning Zhou}
\author[11]{Xiaopeng Zhou}

\affil[1]{School of Physics and Astronomy, Shanghai Jiao Tong University, MOE Key Laboratory for Particle Astrophysics 
and Cosmology, Shanghai Key Laboratory for Particle Physics and Cosmology, Shanghai 200240, China}
\affil[2]{Shanghai Institute of Applied Physics, Chinese Academy of Sciences, Shanghai 201800, China}
%\affil[3]{University of Chinese Academy of Sciences, Beijing 100049, China}
\affil[3]{Key Laboratory of Nuclear Physics and Ion-beam Application (MOE), Institute of Modern Physics, Fudan University, Shanghai 200433, China}
\affil[4]{Shanghai Jiao Tong University Sichuan Research Institute, Chengdu 610213, China}
\affil[5]{Yalong River Hydropower Development Company, Ltd., 288 Shuanglin Road, Chengdu 610051, China}
\affil[6]{School of Physics, Sun Yat-Sen University, Guangzhou 510275, China}
\affil[7]{Tsung-Dao Lee Institute, Shanghai 200240, China}
\affil[8]{School of Physics, Nankai University, Tianjin 300071, China}
\affil[9]{Key Laboratory of Nuclear Physics and Ion-beam Application (MOE), Institute of Modern Physics, Fudan University, Shanghai 200433, China}
\affil[10]{School of Physics, Peking University, Beijing 100871, China}
\affil[11]{School of Physics, Beihang University, Beijing 100191, China}
\affil[12]{International Research Center for Nuclei and Particles in the Cosmos \& Beijing Key Laboratory of Advanced Nuclear Materials and Physics, Beihang University, Beijing 100191, China}
\affil[13]{School of Medical Instrument and Food Engineering, University of Shanghai for Science and Technology, Shanghai 200093, China}
\affil[14]{Department of Physics, University of Maryland, College Park, Maryland 20742, USA}
\affil[15]{School of Mechanical Engineering, Shanghai Jiao Tong University, Shanghai 200240, China}
\affil[16]{School of Physics and Key Laboratory of Particle Physics and Particle Irradiation (MOE), Shandong University, Jinan 250100, China}
\affil[17]{Shanghai Advanced Research Institute, Chinese Academy of Sciences, Shanghai 201210, China}
\affil[18]{Center for High Energy Physics, Peking University, Beijing 100871, China}
\affil[19]{Key Laboratory of Nuclear Physics and Ion-beam Application (MOE), Institute of Modern Physics, Fudan University, Shanghai 200433, China}
\affil[ ]{(PandaX-II Collaboration)}

\date{Jan. 2021}

\maketitle
 
%\begin{document}
\abstract{
We report a systematic determination of the responses of PandaX-II, 
a dual phase xenon time projection chamber detector, to low energy recoils.
The electron recoil (ER) and nuclear recoil (NR) responses are calibrated, respectively, with 
injected tritiated methane or $^{220}$Rn source, and with $^{241}$Am-Be neutron source, within an energy 
range from $1-25$~keV (ER) and $4-80$~keV (NR), under the two drift fields of 400 and 317 V/cm.
An empirical model is used to fit the light yield and charge yield for both types of recoils. The best 
fit models can well describe the calibration data. The systematic uncertainties of the fitted models are obtained
via statistical comparison against the data.
%The general methodology developed here can be adopted in the {\it in situ} calibration of a noble gas TPC. 
}


\section{Introduction}
The nature of dark matter (DM) remains to be one of the most intriguing physics questions today. 
The direct search for an important class of dark matter candidate, the weakly interacting massive particles (WIMPs), 
has been accelerated by the development in dual phase xenon time projection chambers (TPCs), such as PandaX-II~\cite{Tan:2016diz}, XENON-1T~\cite{Aprile:2017aty}, and LUX~\cite{PhysRevLett.112.091303}. In these detectors, a WIMP may interact with xenon nuclei via elastic scattering, %resulting in a few tens of keV$_{\rm nr}$ 
depositing a nuclear recoil (NR) energy from few keV$_{\rm nr}$ to a few tens of keV$_{\rm nr}$. 
%Radioactivities due to (modified as suggested by Qing)
$\gamma$s or $\beta$s  from internal impurities and detector materials produce electron 
recoil (ER) background events, 
%which could be misidentified as the NR signals. (modified as suggested by Qing)
which have a small probability to be identified as the NR signals in these detectors.
%scattering with electrons outside of Xenon nuclei,called electron recoils, are the dominant background in such experiments.

In a dual phase xenon TPC bounded by a cathode at the bottom in the liquid and an anode at the top in the gas, 
each energy deposition will be converted into two channels, the scintillation photons and 
ionized electrons. The former is the so-called $S1$ signal. Electrons are subsequently drifted towards the liquid surface, 
and extracted into the gas region with delayed electroluminescence photons ($S2$) produced. 
Both $S1$s and $S2$s are collected by two arrays of photomultiplier tubes (PMTs) located at the top and bottom of the TPC. 
For a given event, the combination of $S1$ and $S2$ allows the reconstruction of the recoil energy and vertex, 
and the proportion of $S1$ and $S2$ serves as a key discriminant for ER and NR. It is essential to determine the detector response via {\it in situ} calibration.

%It is essential to determine the response
%The key topic of such TPCs is to describe the low energy response from detector calibration. 
%As the good self-shielding of LXe,  external calibration sources are not suitable for large scale Xe detectors, specially for beta and gamma.
For the ER response, several injected sources were
used in PandaX-II, including tritiated methane (CH$_{3}$T), $^{220}$Rn, and $^{83\rm m}$Kr. For NR calibration, 
an external $^{241}$Am-Be (AmBe) neutron source was used. 
In this paper, the detector responses are determined by fitting these data under the NEST2.0~\cite{NEST2p0} prescription.

%calibration data is 
%The data distribution is fitted under the NEST2.0~\cite{xxx} framework, which essentially parameterizes
%the light yield (LY), the charge yield (CY), and their fluctuations as a function of deposited energy.
%Once determined, this model can be used in producing signal and background distributions 
%to fit the dark matter search data.
%A statistical procedure is developed to define the parameter space allowed by the calibration data.

%Thus obtained model provides a self-consistent model 
%in the energy region of interest for dark matter searches. 
%We develop an algorithm based on unbinned log likelihood function to define the allowable region from all the parameters sampling spaces.

This rest of this paper is organized as follows. In Sec.~\ref{sec:calib}, the detector conditions and calibration
setups are introduced. In Sec.~\ref{sec:data_select}, data processing and event 
selection cuts are presented. The response model simulation will be introduced
in Sec.~\ref{sec:sig_model}, followed by detailed discussions on the fits 
of the light yield and charge yield, before the conclusion in Sec.~\ref{sec:conclusion}.

%Section IV 
%details simulations using newly NEST2.0 and interpret our results in terms of light yield and charge yield  Sec. V, summarizes 

\section{Calibration setup}
\label{sec:calib}
% Maybe you can separate the setup part from the history/data part. The history/data part could include the live time, drift field...
The PandaX-II experiment, located at the China Jinping underground laboratory (CJPL)~\cite{Yu-Cheng:2013iaa}, 
was under operation from March 2016 to July 2019, with a total exposure of 132 ton$\cdot$day for 
dark matter search. The operation was divided into three
runs, Runs 9, 10, and 11~\cite{Wang:2020coa}, during which calibration runs were interleaved.
The detector contained 580-kg liquid xenon in its sensitive volume. 
The liquid xenon was continuously purified through two circulation loops, each connected to a 
getter purifier. The internal ER sources were injected through one of the loop.
Two PTFE tubes, at 1/4 and 3/4 height of the TPC surrounding the inner cryostat, were used as the guide tube 
for the external AmBe source. 
The TPC drift field in Run 9 was 400~V/cm and 317~V/cm in Runs 10/11, 
corresponding to a maximum drift time of 350~$\mu$s and 360~$\mu$s, respectively.
The running conditions, key detector parameters, and event selection ranges 
for the calibration data sets are summarized in Table.~\ref{tab:datasets}. 

\begin{table}[!hbtp]
\small
  \centering
  \begin{tabular}{ccccc}\\\hline
  Data set  & Run9 AmBe & Run9 Tritium & Runs 10/11 AmBe & Runs 10/11 $^{220}$Rn\\ \hline
  PDE        &  \multicolumn{2}{c}{ 0.115$\pm$0.002 }        & \multicolumn{2}{c}{0.120$\pm$0.005}            \\ \hline
  EEE        &\multicolumn{2}{c}{0.463$\pm$0.014}    &\multicolumn{2}{c}{0.475$\pm$0.020}\\ \hline
  SEG        &\multicolumn{2}{c}{24.4$\pm$0.4}      &\multicolumn{2}{c}{23.5 $\pm$0.8}\\ \hline
  $E_{\rm{drift}}$ (V/cm) &\multicolumn{2}{c}{400}  &\multicolumn{2}{c}{317}\\ \hline
  $E_{\rm{extract}}$ (kV/cm)&\multicolumn{4}{c}{4.56}\\ \hline
  Duration (day)     & 6.7 & 27.9          &48.5              &11.9\\ \hline
  Number of events & 2902 &9387 &11196 &8841\\ \hline
   Drift time cut($\mu s$) & 18-200 & 18-310& 50-200& 50-350\\ \hline
   range cut & \tabincell{c}{S1:3-150 PE \\  S2:100-20000 PE} & $E_{\rm {rec}}$<25 keV  &\tabincell{c}{S1:3-150 PE \\S2:100-20000 PE} & $E_{\rm {rec}}$<25 keV \\
  \hline
  \end{tabular}
  \caption{Summary of ER and NR calibration data sets and corresponding detector configurations. PDE, EEE, and SEG, respectively, are the photon detection efficiency, electron extraction efficiency, and single electron gain. 
  $E_{\rm drift}$ and $E_{\rm extract}$ are the drift field and extraction field. 
  The number of events correspond to the calibration 
  data after all cuts, which are described in Sec.~\ref{sec:data_select}.
  }
  \label{tab:datasets}
\end{table}

\subsection{Tritiated methane} 
\begin{figure}[!htbp]
\centering
\includegraphics[scale=0.3]{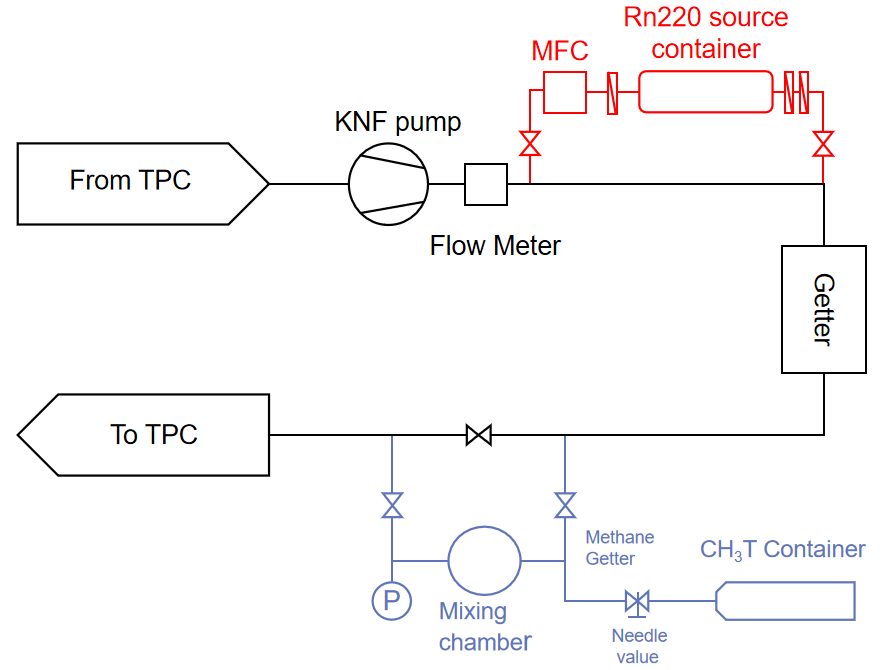}
\caption{Tritiated methane (blue) and $^{220}$Rn (red) injection system.}
\label{fig:injection diagram}
\end{figure}

Tritiated methane calibration was first developed in the LUX experiment~\cite{Akerib:2015wdi}, which provided excellent 
internal low energy $\beta$ events.
The tritiated methane source used in PandaX-II was procured from 
American Radio labeled Chemicals, Inc., with a specific activity of 0.1~Curie per mole of methane. 
The injection diagram is shown in Fig.~\ref{fig:injection diagram}.
The tritiated methane bottle was immersed in a liquid-nitrogen cold trap, so that controllable amount of 
the CH$_{3}$T gas could diffuse through a needle valve 
to the 100 mL mixing volume. The gas in the mixing volume was flushed with xenon gas into the detector.

The injection of tritium was performed in 2016 right after Run 9, during which 
about 5.4$\times 10^{-10}$~mol of methane was loaded into the detector.
The tritium events were distributed uniformly in the detector. 
Liquid xenon was
constantly circulated at a speed of about 40 SLPM (standard
liter of gas per minute) through the purifier. The entire calibration run lasted for 44 days, 
and the later data set with an average electron lifetime of 706~$\mu$s is used as the ER
calibration data.

It was realized that the hot getters were inefficient to remove tritium, whose activity plateaued at 
10.2~$\mu$Bq/kg. A distillation campaign was carried out after the calibration, which reduced the tritium 
activity to 0.049$\pm$0.005~$\mu$Bq/kg in Runs 10/11~\cite{Wang:2020coa}.

\subsection{$^{220}$Rn}
$^{220}$Rn, a decay progeny of $^{232}$Th, is a naturally occurring radioactive noble gas isotope. With a half-life of 55~s, 
it poses much less risk to contaminate the liquid 
xenon TPC, as first demonstrated in XENON100~\cite{Aprile:2016pmc}.
The details of $^{220}$Rn calibration setup and operation in PandaX-II can be found in Ref.~\cite{Ma:2020kll}.
The injection system consisting of a mass flow controller and a $^{232}$Th source chamber 
with filters upstream and downstream, is shown in Fig.~\ref{fig:injection diagram}. 
After $^{220}$Rn was injected into the detector, the $\beta$-decay of the daughter nucleus $^{212}$Pb 
gives uniformly distributed ER events with an energy extending to zero. 11.9~days of $^{220}$Rn data in 2018 
are used as the low energy ER calibration for Runs 10/11.

%which uniformly distributed in the detector as the long half-life, 10.6h, provides it sufficiently long time to spread throughout the entire detector volume. At the same time, this half-life is sufficiently short to allow the activity to decay within a week. 
%We did three $^{220}$Rn injection operation in 2017, 2018 and 2019 with different thorium source.
 %In this injection, lantern mantles treated with thorium nitrate (Th(NO$_{3}$)$_{4}$) were used as the radon sources. 
%A high energy veto trigger of DAQ system was implemented, which reject events with more than 70 hit PMTs.  

\subsection{AmBe}
Neutron calibration data with an AmBe ($\alpha$, n) source~\cite{Wang:2019opt} were taken during Run 9 and Runs 10/11.
The source was placed inside the external calibration tubes.
Calibration runs were taken at eight symmetric locations in each loop to evenly sample the detector.
For different source locations, no significant difference is identified in the detector response, 
so the data are grouped together in the analysis.

\section{Data selection}
\label{sec:data_select}

The processing of the calibration data follows the procedure in Ref.~\cite{Wang:2020coa}. Compared to previous analyses~\cite{Tan:2016diz,Xiao:2015psa}, seven unstable PMTs are inhibited from all data sets for consistency. Improvements are made on the PMT gain calibration, quality cuts, position reconstruction, and corresponding non-uniformity correction. 

The raw $S1$ and $S2$ of each event have to be first corrected for position non-uniformity, 
based on the three-dimensional variation of the raw $S1$ and $S2$ for the uniformly distributed mono-energetic events, 
e.g. 164 keV ($^{131\rm m}$Xe) due to activation from the neutron source. The correction to $S1$ is a smooth three-dimensional 
hyper-surface. The correction to $S2$ is separated into an exponential attenuation vs. drift time 
(electron lifetime $\tau$), and a 
smooth two-dimensional surface in the horizontal plane.

The electron equivalent energy of each event is reconstructed as 
\begin{equation}
\label{eq:energy_comb}
  E_{\rm{rec}} = W \times
  \left(\frac{S1}{\rm{PDE}}+\frac{S2}{\rm{EEE}\times\rm{SEG}}\right)
\end{equation}
where $W=13.7$~eV~\cite{Szydagis:2011tk} is the average energy to produce either a scintillation photon or free electron in liquid xenon, and
PDE, EEE, and SEG, respectively, are the photon detection efficiency (ratio of detected photoelectrons to the total photons), electron extraction efficiency, and single electron gain, obtained from the data (see Ref.~\cite{Wang:2020coa} and Table.~\ref{tab:datasets}). 

\begin{figure}[!htbp]
  \centering
  \begin{subfigure}{0.4\textwidth}
    \includegraphics[width=1.0\textwidth]{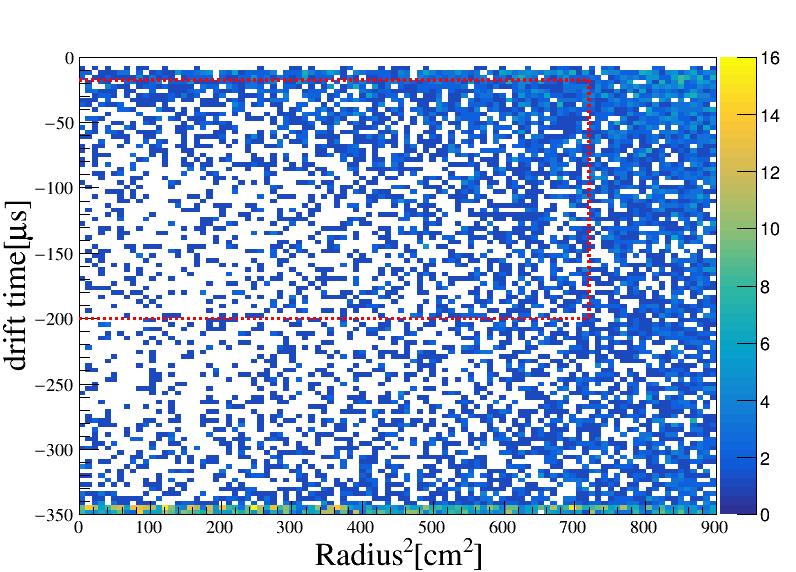}
    \caption{Run 9 AmBe}
    \label{fig:run9 ambe vertex}
  \end{subfigure}
  \begin{subfigure}{0.4\textwidth}
    \includegraphics[width=1.0\textwidth]{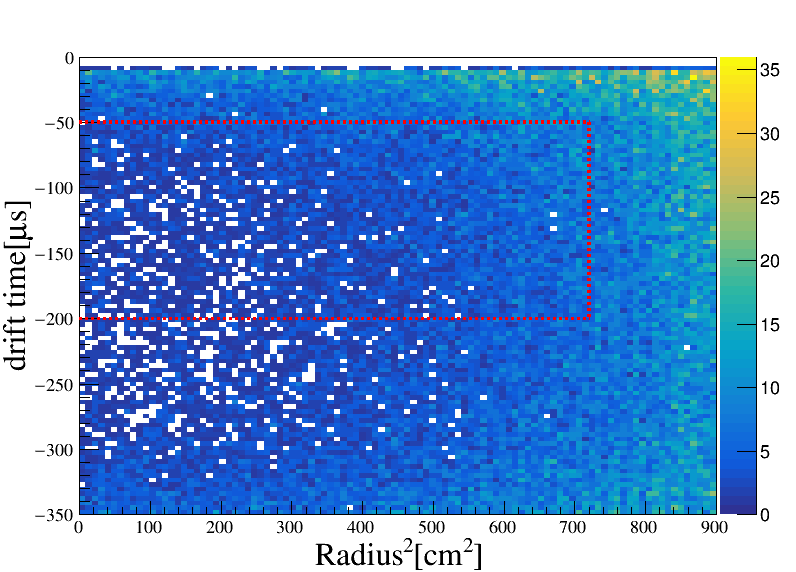}
    \caption{Runs 10/11 AmBe}
    \label{fig:run11 ambe vertex}
  \end{subfigure}
  \begin{subfigure}{0.4\textwidth}
    \includegraphics[width=1.0\textwidth]{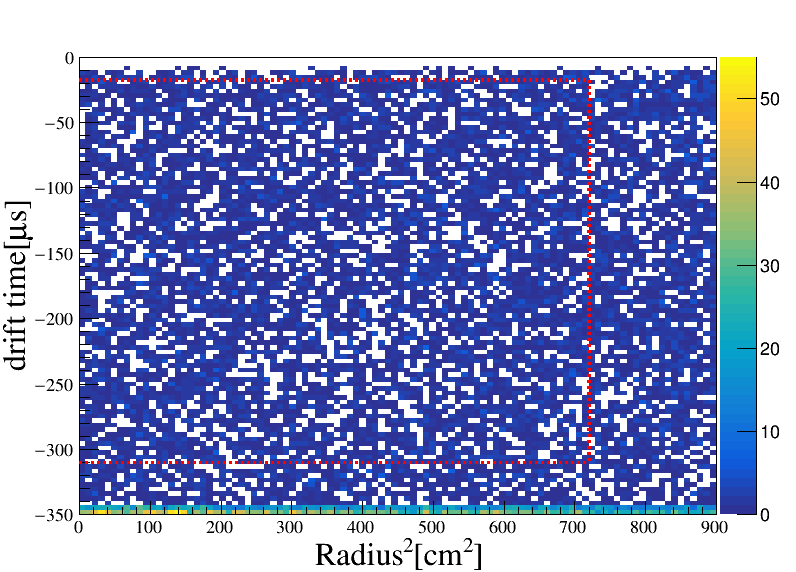}
    \caption{Run 9 tritium}
    \label{fig:run9 tritium vertex}
  \end{subfigure}
  \begin{subfigure}{0.4\textwidth}
    \includegraphics[width=1.0\textwidth]{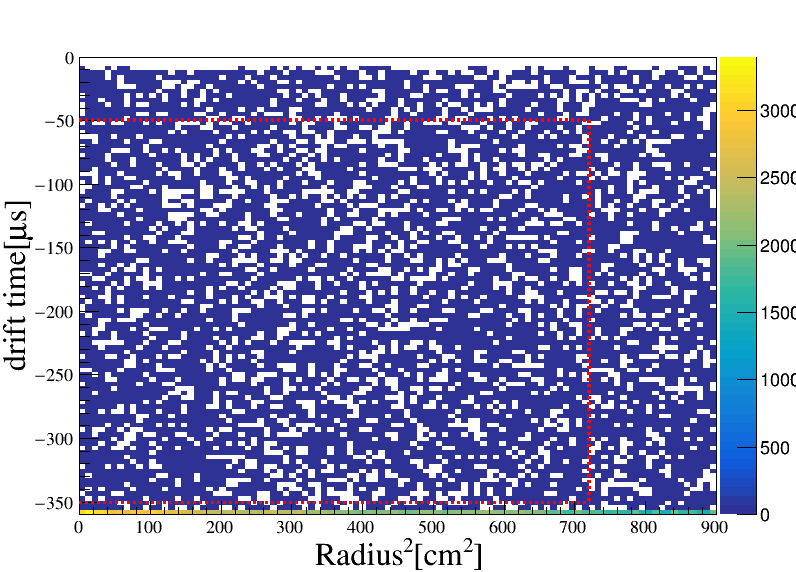}
    \caption{Runs 10/11 Radon}
    \label{fig:run11 tritium vertex}
  \end{subfigure}
  \caption{Event vertex distribution in drift time vs. radius-squared for each calibration data set. The FV region is indicated by dashed red line in each figure.}
  \label{fig:vertex distribution}
\end{figure}

\begin{figure}[!htbp]
  \centering
   \begin{subfigure}{0.4\textwidth}
    \includegraphics[width=1.0\textwidth]{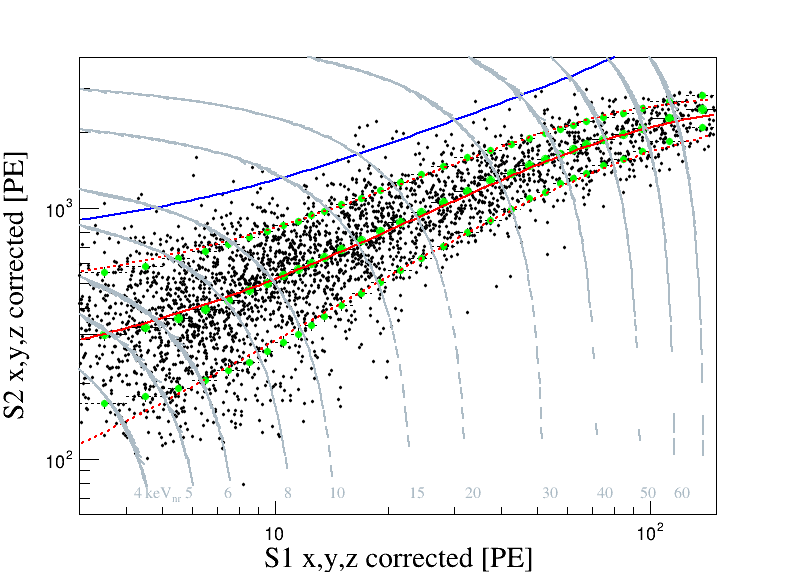}
    \caption{AmBe in Run9}
    \label{fig:run9 ambe band median}
  \end{subfigure}
  \begin{subfigure}{0.4\textwidth}
    \includegraphics[width=1.0\textwidth]{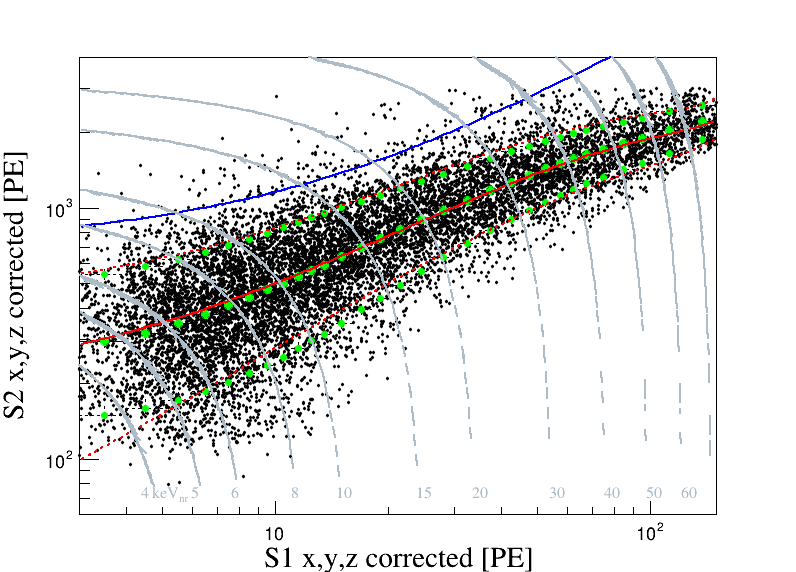}
    \caption{AmBe in Runs 10/11}
    \label{fig:run11 ambe band median}
  \end{subfigure}
  \begin{subfigure}{0.4\textwidth}
  \includegraphics[width=1.0\textwidth]{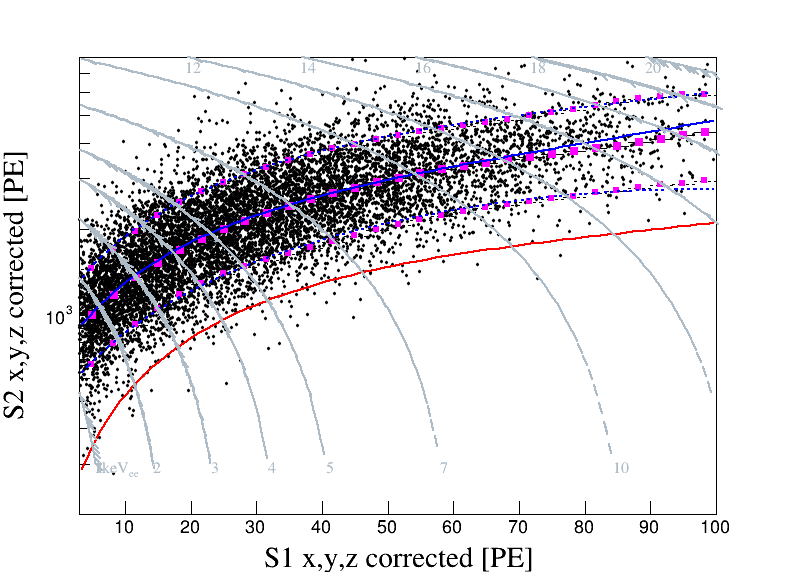}  
    \caption{Tritium in Run 9}
    \label{fig:run9 tritium band median}
  \end{subfigure}
  \begin{subfigure}{0.4\textwidth}
    \includegraphics[width=1.0\textwidth]{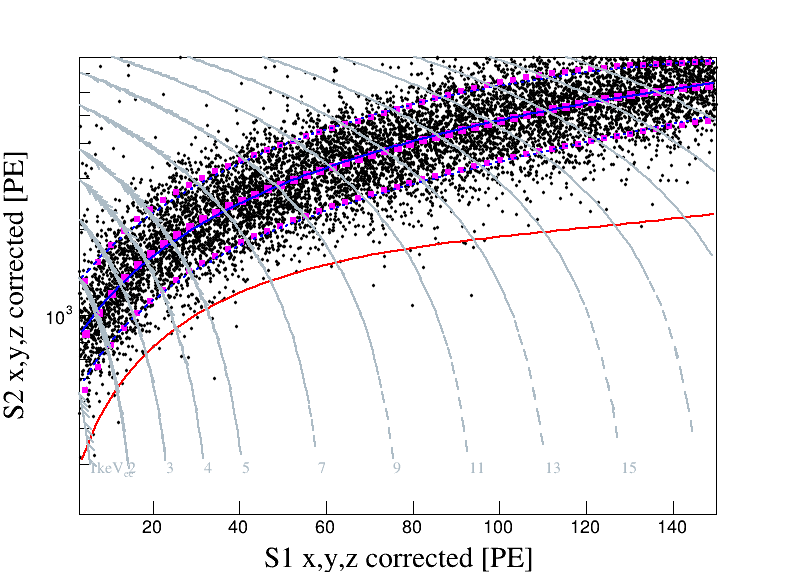}
    \caption{Radon in Runs 10/11}
    \label{fig:run11 radon band median}
  \end{subfigure}
  \caption{
  $S2$ vs. $S1$ of the selected calibration events. The red (blue) solid lines are the medians of NR (ER), and the red (blue) dashed lines refer to the 90\% quantiles. For comparison, the 90\% quantiles from the best fit response models (Sec.~\ref{sec:sig_model}) are overlaid as the green (purple) dotted lines for NR (ER). The gray dashed 
  curves are the equal-$E_{\rm nr}$ and equal-$E_{\rm rec}$ lines for the NR and ER events, respectively. 
  %The black dots are data points in log10(S2/S1) versus S1. The red solid and dashed lines are the median and 2-sigma quantile of data. The green ones are that of simulation. The gray dashed curves are the equal energy curves with NR energy and ER energy indicated in the figures.
  }
  \label{fig:2D band}
\end{figure}

% selection window for calibration: S1, S2 range, FV
Events with a single pair of $S1$ and $S2$ are chosen. Fiducial volume (FV) definition is consistent with Ref.~\cite{Wang:2020coa}, except that a lower 
cut in drift time (200~${\mu}$s) is applied to the AmBe data to avoid events that multi-scatter and 
deposit part of the energy in the below-cathode region, leading to suppressed $S2$~\cite{Xiao:2015psa}. 
The lower selection cuts $S1>3$~PE and $S2_{\rm raw}>100$~PE are applied to all data sets.
For the AmBe data, the upper selection cut is set at $S1<150$~PE ($\sim$80~keV$_{\rm nr}$). 
For ER data, events with $E_{\rm rec}<25$ keV are selected. 
The vertex distributions of selected events are shown in Fig.~\ref{fig:vertex distribution}, with FV cuts indicated.
The distributions of $S2$ vs. $S1$ for ER and NR events are shown in Fig.~\ref{fig:2D band}, 
which will be used to determine the detector response model.
% how many events are selected in this paper
% log10(S2/S1) vs. S1 band ???
% vertex distribution figure may be deleted

%\section{Analysis}
\section{Determination of PandaX-II response models}
\label{sec:sig_model}
Our ER and NR response models follow the prescription of NEST2.0~\cite{NEST2p0}, but use our own customization.
The light yield ($L_y$) and charge yield ($Q_y$), defined as the number of initial quanta (photons or ionized electrons) per unit recoil energy, can be parameterized and fitted to the calibration data. We shall discuss the simulation models in Sec.~\ref{sec:quanta_model} and Sec.~\ref{sec:det_model}, which will be then fitted to data in Sec.~\ref{sec:analysis_method}

\subsection{Quanta generation}
\label{sec:quanta_model}
\begin{figure}[!htbp]
  \centering
   \begin{subfigure}{0.4\textwidth}
    \includegraphics[width=1.0\textwidth]{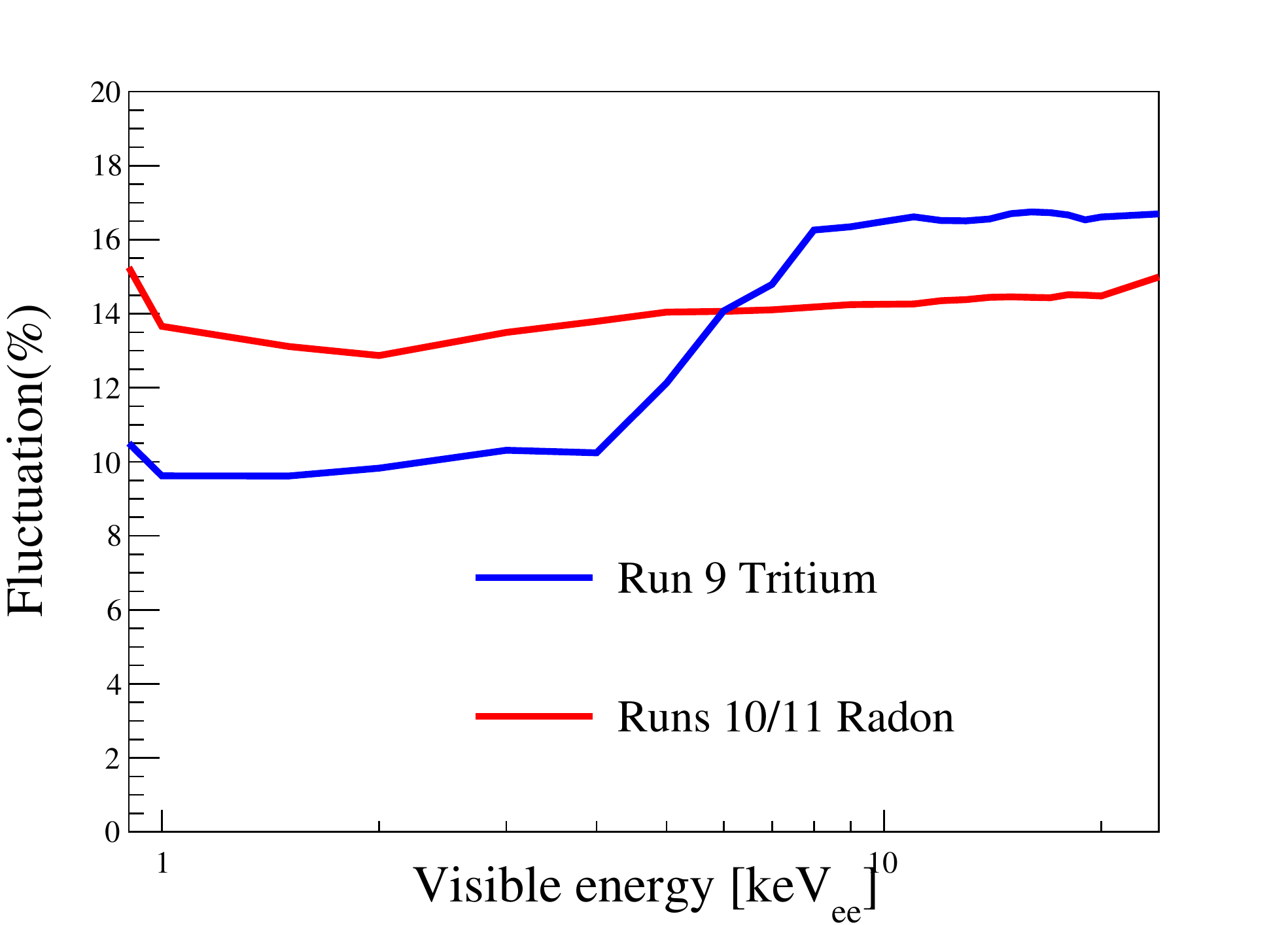}
    \caption{ER}
    \label{fig:er fluctuation}
  \end{subfigure}
  \begin{subfigure}{0.4\textwidth}
    \includegraphics[width=1.0\textwidth]{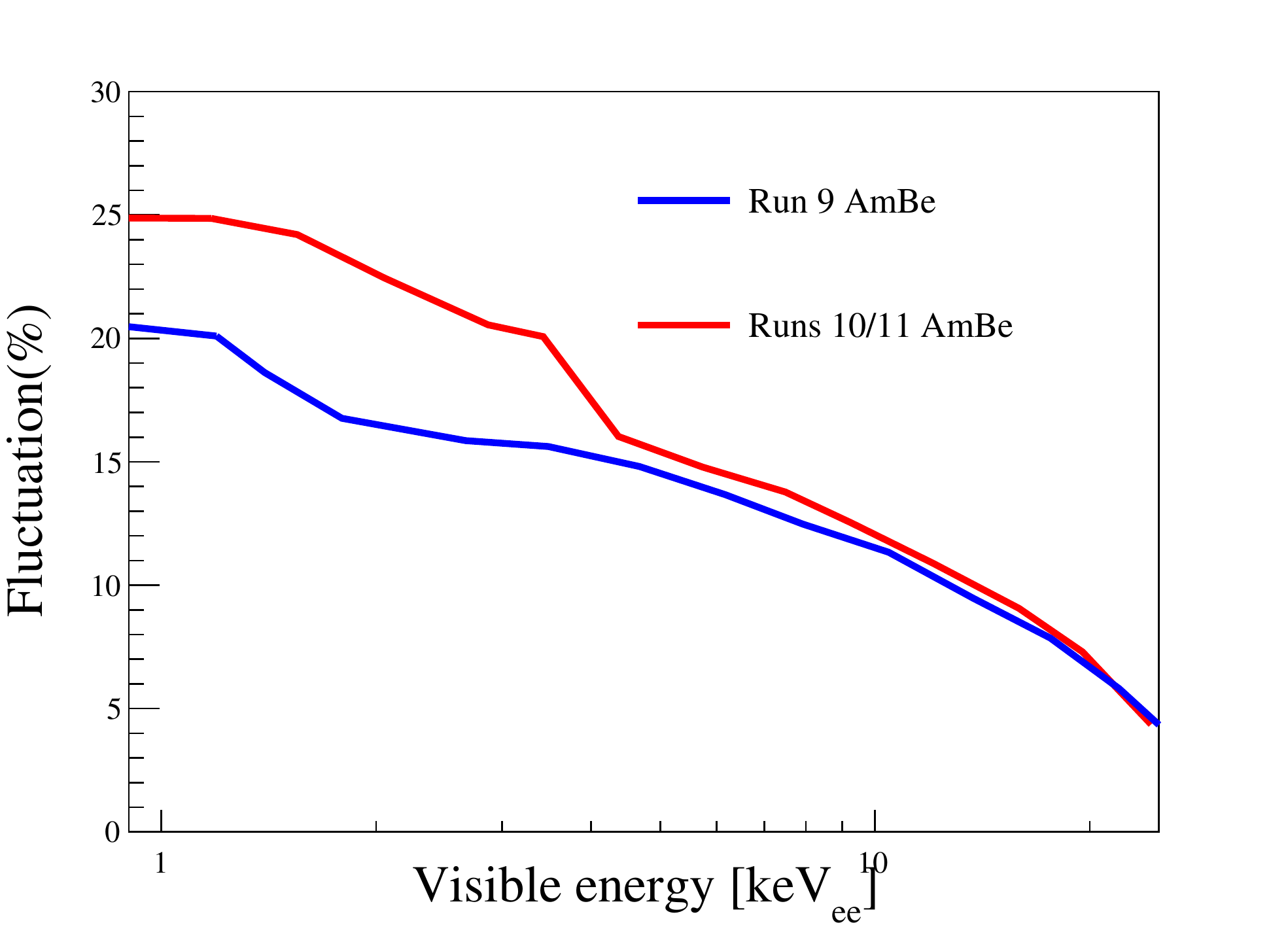}
    \caption{NR}
    \label{fig:nr fluctuation}
  \end{subfigure}
  \caption{The empirical fluctuation parameter $f$ as a function visible energy for the ER (left) and NR (right) data, both in electron equivalent energy keV$_{\rm {ee}}$}
  \label{fig:Ne_fluc}
\end{figure}

For a distribution of true recoil energy from the calibration source, each recoil energy $E_0$ is converted into two types of quanta, scintillation photons $n_{\rm ph}^0$ or ionized electrons $n_{\rm e}^0$. For the NRs, the visible energy is quenched into $E_0\times L$
due to unmeasurable dissipation of heat in the recoil, 
where $L$ is the so-called Linhard factor with a value ranging from 0.1 to 0.25 for 
$E_0$ less than 100~keV$_{\rm nr}$~\cite{NEST1.0NR}.
For ER events, on the other hand, $E_0$ converts almost entirely to photons or electrons, so effectively $L=1$. 
Now the number of quanta can be expressed as
\begin{equation}
\label{equation:quanta}
\begin{aligned}
n_{\rm q} &\equiv n_{\rm ph}^0+n_{\rm e}^0 = \frac{E_{0} L}{W} \\ {n_{\rm ph}^0} & = L_y E_0\,, {n_{\rm e}^0} = Q_y E_0 \end{aligned}
\end{equation}
 In NEST2.0, $L_y$ is parameterized as an empirical function of $E_0L$, and $L_y$ and $Q_y$ are connected through Eqn.~\ref{equation:quanta}.
The intrinsic (correlated) fluctuations in ${n_{\rm e}^0}$ and ${n_{\rm ph}^0}$ is encoded in our simulation by an energy 
dependent Gaussian smearing function $f(E_{0}L)$ as
\begin{equation}
\label{equation:fluctuation}
\begin{aligned}
n_{\rm e} &= {\rm Gaus}({n_{\rm e}^0} , f(E_{0}L)\times {n_{\rm e}^0}) \\
n_{\rm ph} &= n_{\rm q} - n_{\rm e} \,, 
\end{aligned}
\end{equation}
in which ${\rm Gaus}(\mu,\sigma)$ is a Gaussian random distribution with $\mu$ and $\sigma$ as the mean and 1$\sigma$ value, and $f$ can be adjusted to the data (see Fig.~\ref{fig:Ne_fluc}).

%The photons are detected as $S1$. The electrons drift upward to the gas region and detected as $S2$. 
%\color{blue}
%We tried another simpler process, in which we calculate the $n_{ph}$ and $n_e$ directly by $LY\times E$ and $CY\times E$. An overall fluctuation is applied. In this scenario, we can get a quick model to describe data. 
%\color{black}

\subsection{Model of the detector}
\label{sec:det_model}
The detector model is used to convert $n_{\rm ph}$ and $n_{\rm e}$ to detected $S1$ and $S2$. For the R11410-20 PMTs used in PandaX-II, 
the double-photoelectron emission probability by the 178 nm scintillation photons $p_{\rm2pe}$ 
is measured to be $0.21\pm0.02$ from the data~\cite{Wang:2020coa}. Therefore, the number of detected photons ($N_{\rm dph}$) can be simulated as
 is
 \begin{equation}
 \begin{aligned}
    {N_{\rm dph}} &= {\rm Binom}(n_{\rm ph}, {\rm PDE}/(1 + {p_{\rm2pe}}))\\
    \end{aligned}
\end{equation}
in which ${\rm Binom}(N,p)$ refers to a binomial distribution with $N$ throws and a probability $p$, and $\rm{PDE}/(1+p_{\rm2pe})$ is the binomial probability to detect a photon.
$N_{\rm dph}$ is randomly distributed onto the two arrays of PMTs (55 each) according to the 
measured top/bottom ratio from the data ($\sim$1:2). Each detected photons are then fluctuated 
by $p_{\rm2pe}$, leading to the number of photoelectrons
\begin{equation}
 \begin{aligned}
{N_{\rm{PE}}} &= {N_{\rm dph}} + {\rm Binom}({N_{\rm dph}}, p_{\rm 2pe})\,.
 \end{aligned}
\end{equation}
%then Gaussian-fluctuated by the single photoelectron (SPE) resolution $\sigma$ of 33\%~\cite{Li:2015qhq},
%leading to the number of photoelectrons
$S1$ can be subsequently obtained by applying the single photoelectron (SPE) resolution, 
modeled as a Gaussian with a $\sigma_{\rm SPE}$ of 33\% ~\cite{Li:2015qhq} 
\begin{equation}
    S1 = {{\rm Gaus}(N_{\rm PE}},\sigma_{\rm SPE} \times \sqrt{N_{\rm PE}})\,.
\end{equation}
Each $S1$ is required to have at least three hits, with each hit larger than 0.5~PE to simulate the single channel readout threshold and the multiplicity cut in the analysis~\cite{Wang:2020coa}.

Similarly, $S2$ is simulated based on $n_{\rm e}$ by using detector parameters from the data. For each event, 
the drift time $t_{\rm drift}$ is randomized according to the data distribution, leading to an electron survival probability $s=\exp(-t_{\rm drift}/\tau)$ with the electron lifetime $\tau$ obtained from the data. So at the liquid level, the number of electrons is
\begin{equation}
    {N_{\rm e}^{'}} = {\rm Binom}(n_{\rm e}, s)\,.
\end{equation}
Then the number of extracted electron $N_{\rm e}^{''}$ and $S2$ can be simulated as
\begin{equation}
\begin{aligned}
    N_{\rm e}^{''} &=  {\rm Binom}(N_{\rm e}^{'}, {\rm EEE})\, ,\\
    S2 &= {\rm Gaus}({N_{\rm e}^{''}} \times {\rm SEG},\sigma_{\rm SE} \times \sqrt{{N_{\rm e}^{''}}})\,,
    \end{aligned}
\end{equation}
in which $\sigma_{\rm SE}\sim 8.3$~PE is the Gaussian width for the single electron signals. 

As discussed in Ref.~\cite{Wang:2020coa}, the nonlinearities in $S1$ and $S2$ due to baseline suppression firmware 
are measured from the data, denoted as $f_1(S1)$ and $f_2(S2)$. So the detected $S1$ and $S2$ are
\begin{equation}
    S1_{d} = S1 \times f_{1}\,, S2_d = S2 \times f_{2}\,.
\end{equation}

Finally, the data selection efficiency is parameterized as a Fermi-Dirac function
\begin{equation}
\label{eq:eff}
\epsilon(S1_d) = \frac{1}{1+\exp(\frac{S1_d-p_0}{p_1})}\,,
\end{equation}
where $p_0$ and $p_1$ will be determined by fitting to calibration data. 
Note that our selection efficiency on $S2$ is expected to be 100$\%$, since the $S2$ selection cut 
is at 100~PE, significantly higher than the hardware trigger efficiency of 50~PE~\cite{Wu:2017cjl}.

\subsection{Extraction of parameters in the response model}
\label{sec:analysis_method}
%\section{Comparison of calibration data and signal model}
% including parameter fluctuation, spectra comparison, efficiency, threshold...
In this section, the ER and NR response models will be fitted against the calibration data in $S1$ and $S2$ using 
unbinned likelihood. The systematic uncertainties of the models
are quantified by a likelihood ratio approach.

%is determined by mock data with comparable statistics 
%to the calibration data. 
%For each set of model parameters, a probability density distribution function (PDF)
%in $S1$ and $S2$ is produced with large number of events, and compared to the data via 
%unbinned likelihood, from which the best fit can be obtained. The allowable parameter space, 
%quantified by the likelihood, is determined by mock data with comparable statistics 
%to the calibration data. 

\subsubsection{The likelihood function}
As an initial approximation, 
$L_y$ can be fitted from the medians of the calibration data distribution as
 \begin{equation}
\label{equation:data yield}
L_y^0(E_{\rm rec}/L) = \frac{S1}{{\rm PDE}\times E_{\rm rec}/L}\,,
\end{equation}
where $E_{\rm rec}$ (Eqn.~\ref{eq:energy_comb}) is the reconstructed energy including all detector effects, and the $\frac{\rm{E_{rec}}}{L}$ is the estimate of $E_0$. The true
$L_y$ can be parameterized as 
\begin{equation}
\label{eq:leg_exp}
    L_y(E_0) = L_y^0(E_0) + \sum_{n=0} ^{4} c_n P_n(E_0) \,,
\end{equation}
in which $P_n(E_0)$ is the $n$th order Legendre polynomial functions, and $c_n$ can be fitted to data.

For a given model, a two-dimensional probability density function (PDF) in ($S1$,$S2$) is produced 
with a large statistics simulation described in Secs.~\ref{sec:quanta_model} and~\ref{sec:det_model} 
using the following sets of parameters: 
a) PDE, EEE and SEG constrained by their Gaussian priors (Table~\ref{tab:datasets}), 
with the anti-correlation between PDE and EEE embedded (see Ref.~\cite{Cui:2017nnn}),
b) parameters for $\epsilon(S1_d)$ in Eqn.~\ref{eq:eff}, 
with a flat sampling of $p_0\in(2,5)$ and $p_1\in(0,1)$, respectively,
and c) a 4th order Legendre polynomial expansion for $L_y$ in Eqn.~\ref{eq:leg_exp}, with $c_n$($n=0,1,2,3,4$) uniformly sampled from $-5$ to 5.
Other parameters that are independently determined from the data 
are fixed in the simulation, such as the fluctuation in $n_{\rm e}$, 
the electron lifetime, $p_{\rm2pe}$, and the baseline suppression nonlinearities. 

To compare the data with the PDF, a standard unbinned log likelihood function is defined in the space of ($S1$,$S2$) as
\begin{equation}
%\begin{aligned}
     -2\ln\mathcal{L} = \sum_{i=1}^{N} -2\ln(P(S1^i, S2^i))\,\\
     %t_{\theta} =  \mathcal{L(\theta)} -  \mathcal{L}(\theta_{0})
%    \end{aligned}
\end{equation}
in which $P(S1^i, S2^i)$ is the probability density for a given calibration data point $i$, and $N$ is the total number of events for each calibration data set. 

\subsubsection{The best fit and allowable parameter space}
An independent parameter scan is carried out to determine the best fit model for each calibration data set. The best fit corresponds to the PDF which gives the minimum $-2\ln\mathcal{L}$. For illustration, the centroids and 90\% quantiles of the best fit models from the four data sets are 
overlaid in Fig.~\ref{fig:2D band}, where good agreements with the data are observed.
%The best fit g1 and g2 are 0.1135 and 11.29 for Run 9 and 0.1199 and 11.11 for Runs 10/11 within our nominal values.
%in Table~\ref{tab:datasets}.

%The example sampling in the parameter space of $L_y$ is illustrated in Fig.~\ref{fig:Legendre's polynomial}. 
%\begin{figure}[!htbp]
 % \centering
  %\begin{subfigure}{0.49\textwidth}
  %  \includegraphics[width=1.0\textwidth]{figure/run11_ambe_pol4.pdf}
   % \caption{NR models for Runs 10/11}
    %\label{fig:cy_nr_worlddata}
  %\end{subfigure}
  %\begin{subfigure}{0.49\textwidth}
   % \includegraphics[width=1.0\textwidth]{figure/run11_radon_pol4.pdf}
    %\caption{ER models for Runs 10/11}
    %\label{fig:run11 radon pol4}
  %\end{subfigure}
  %\caption{Example $L_y$ model scan during likelihood fits to data.Each set responds to one set of }
  %\label{fig:Legendre's polynomial}
%\end{figure}

\begin{figure}[!htbp]
  \centering
   \begin{subfigure}{0.4\textwidth}
    \includegraphics[width=1.0\textwidth]{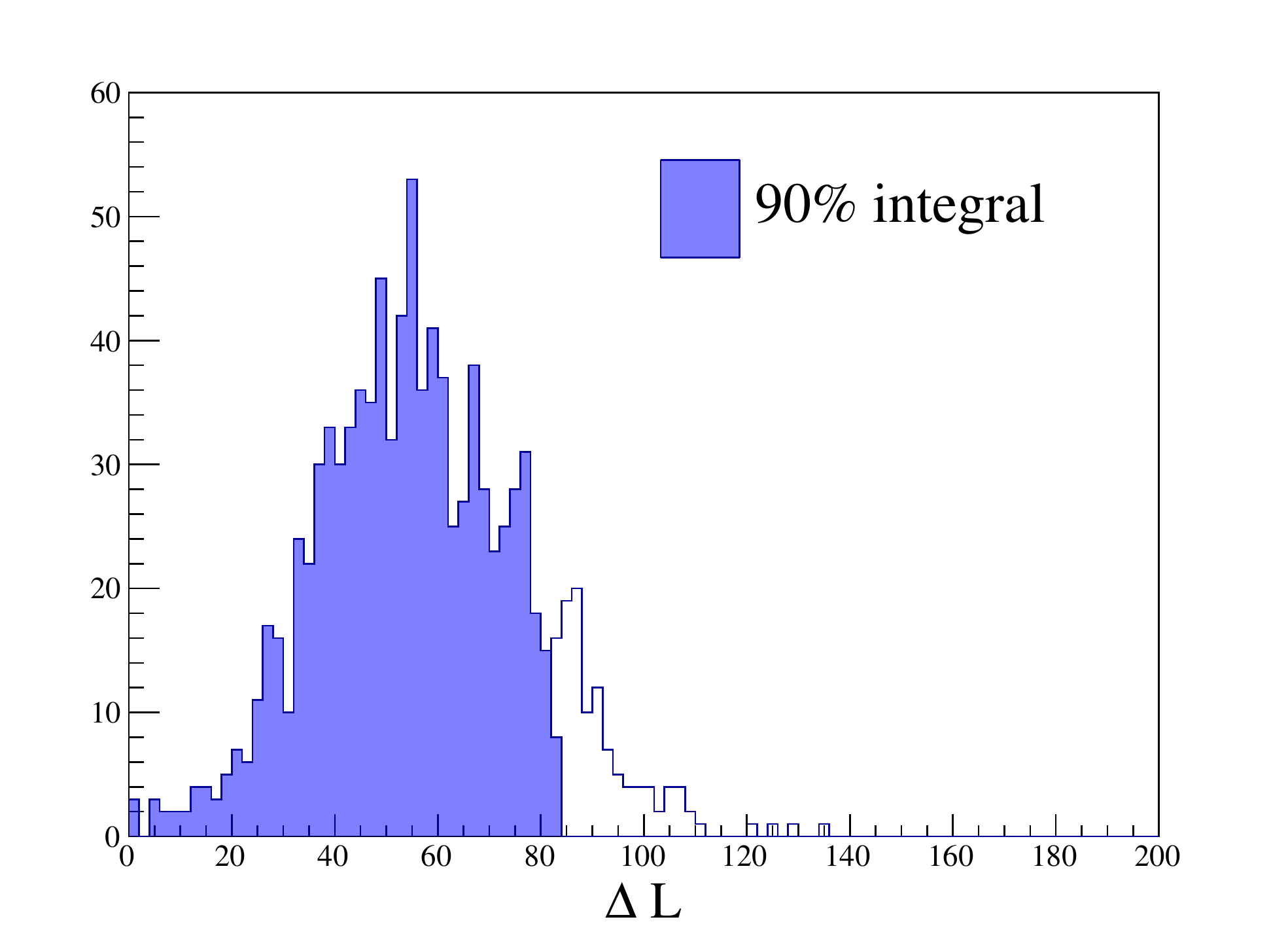}
    \caption{AmBe in Run 9}
    \label{fig:run9 ambe delta L}
  \end{subfigure}
  \begin{subfigure}{0.4\textwidth}
    \includegraphics[width=1.0\textwidth]{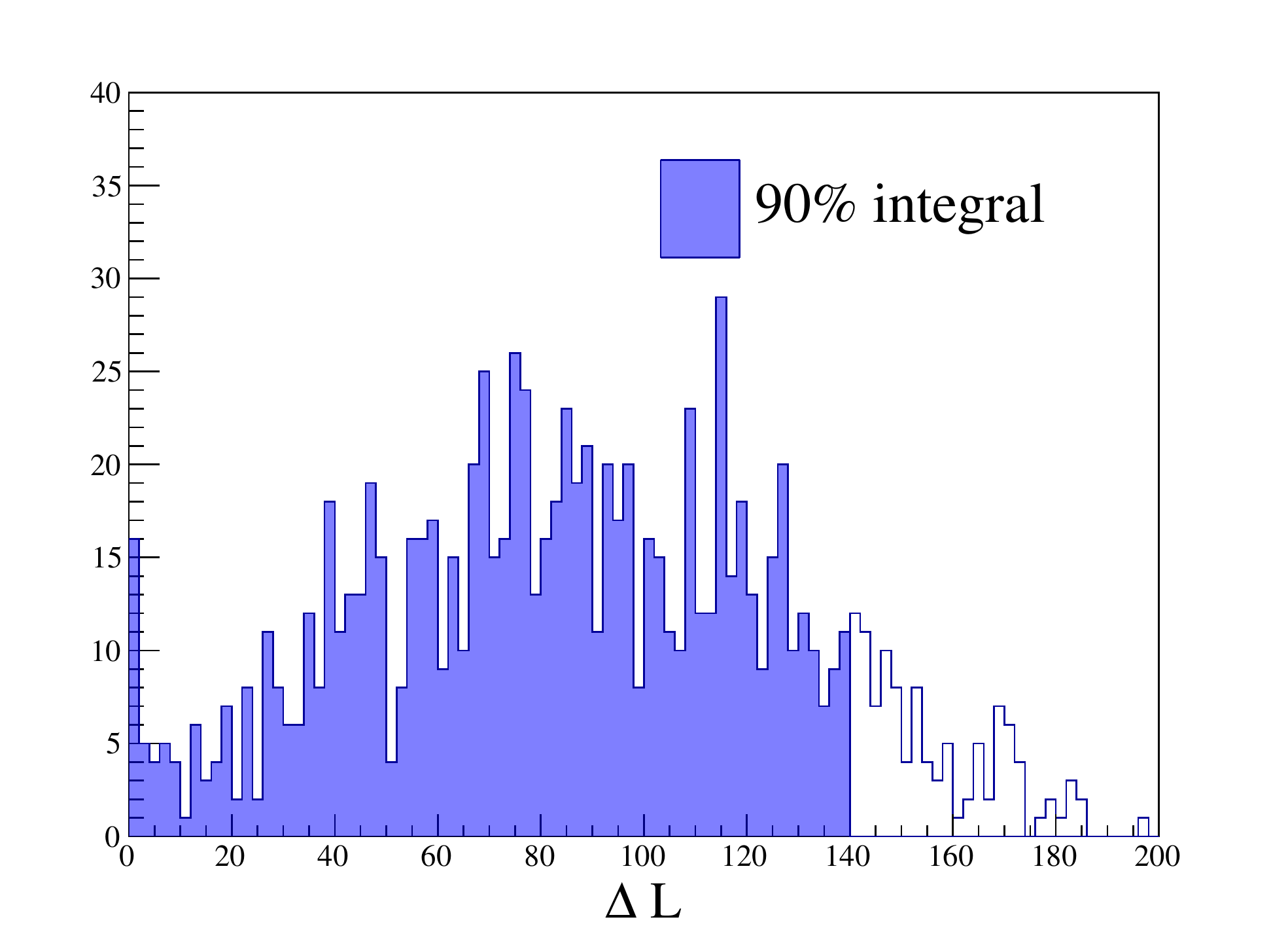}
    \caption{AmBe in Runs 10/11}
    \label{fig:run11 ambe delta L}
  \end{subfigure}
  \begin{subfigure}{0.4\textwidth}
  \includegraphics[width=1.0\textwidth]{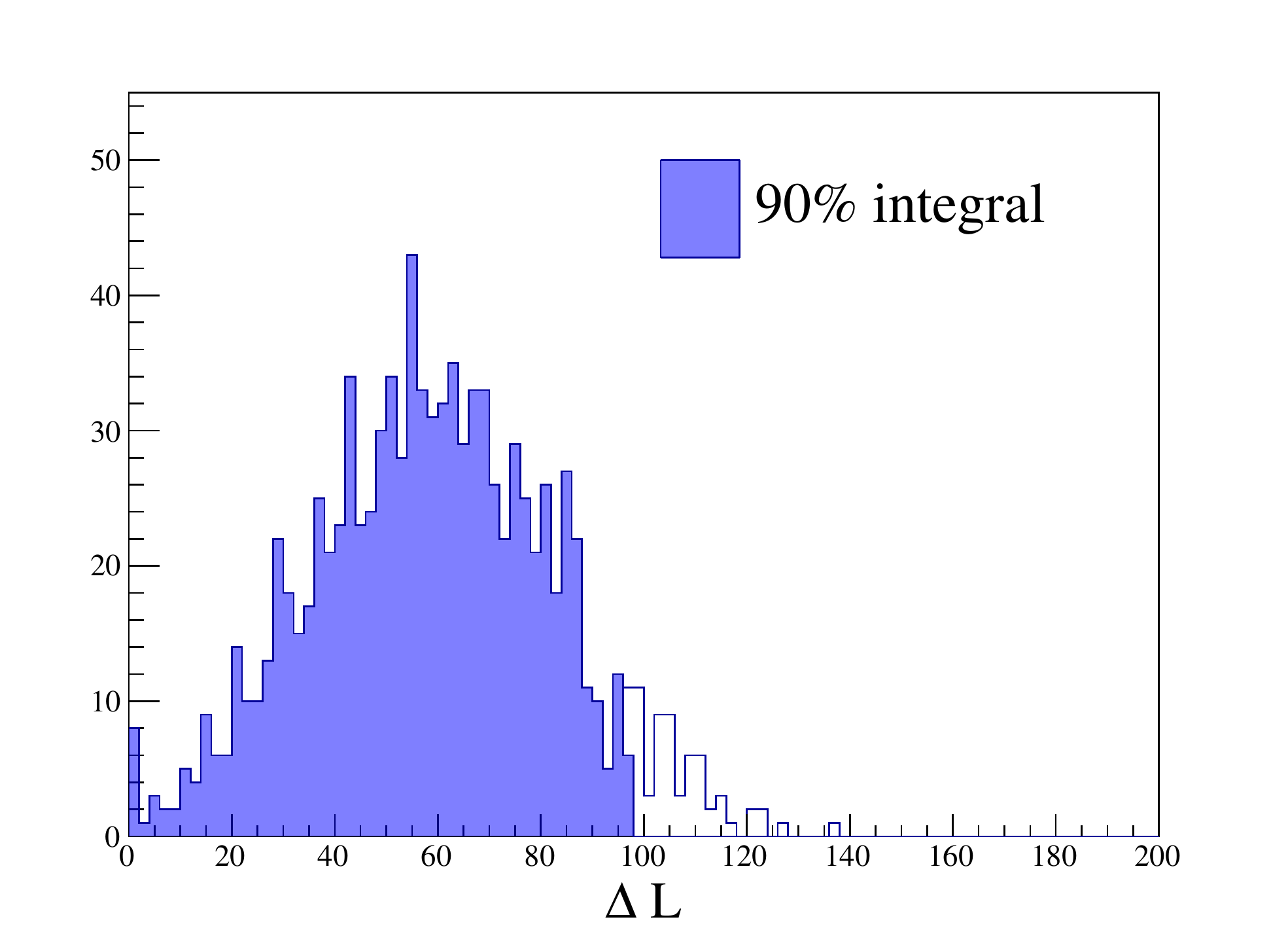} 
    \caption{Tritium in Run 9}
    \label{fig:run9 tritium delta L}
  \end{subfigure}
  \begin{subfigure}{0.4\textwidth}
    \includegraphics[width=1.0\textwidth]{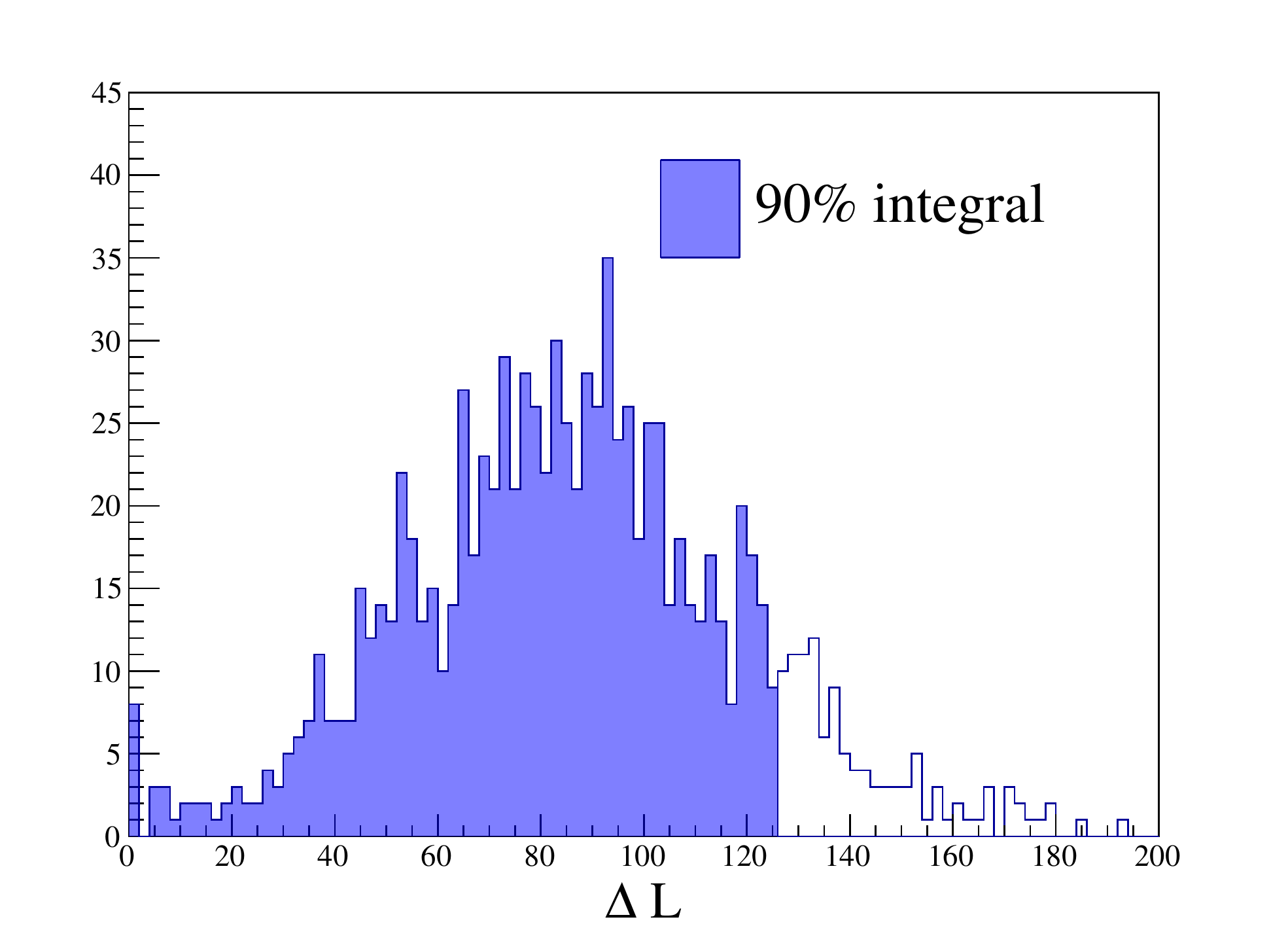}
    \caption{Radon in Runs 10/11}
    \label{fig:run11 radon delta L}
  \end{subfigure}
  \caption{The distribution of $\Delta \mathcal{L}$ for mock calibration data sets generated at 
  the best fit parameter points. The shaded regions indicate the 90\% integrals.}
  \label{fig:acceptance region}
\end{figure}

\begin{figure}[htbp!]
  \centering  
  \begin{subfigure}{0.3\textwidth}
    \includegraphics[width=1\textwidth]{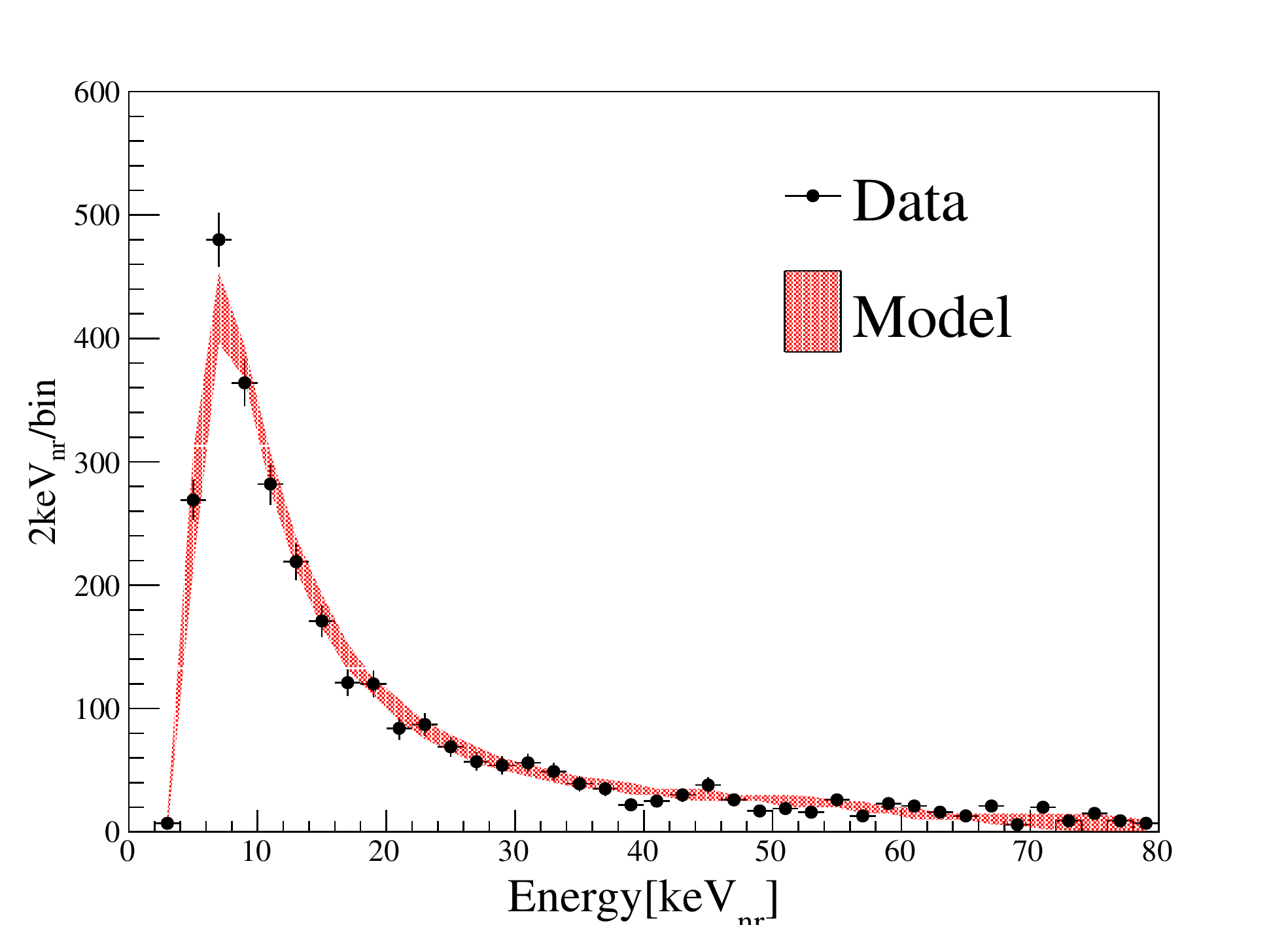}
    \caption{Run 9 AmBe energy}
  \end{subfigure}
  \begin{subfigure}{0.3\textwidth}
    \includegraphics[width=1\textwidth]{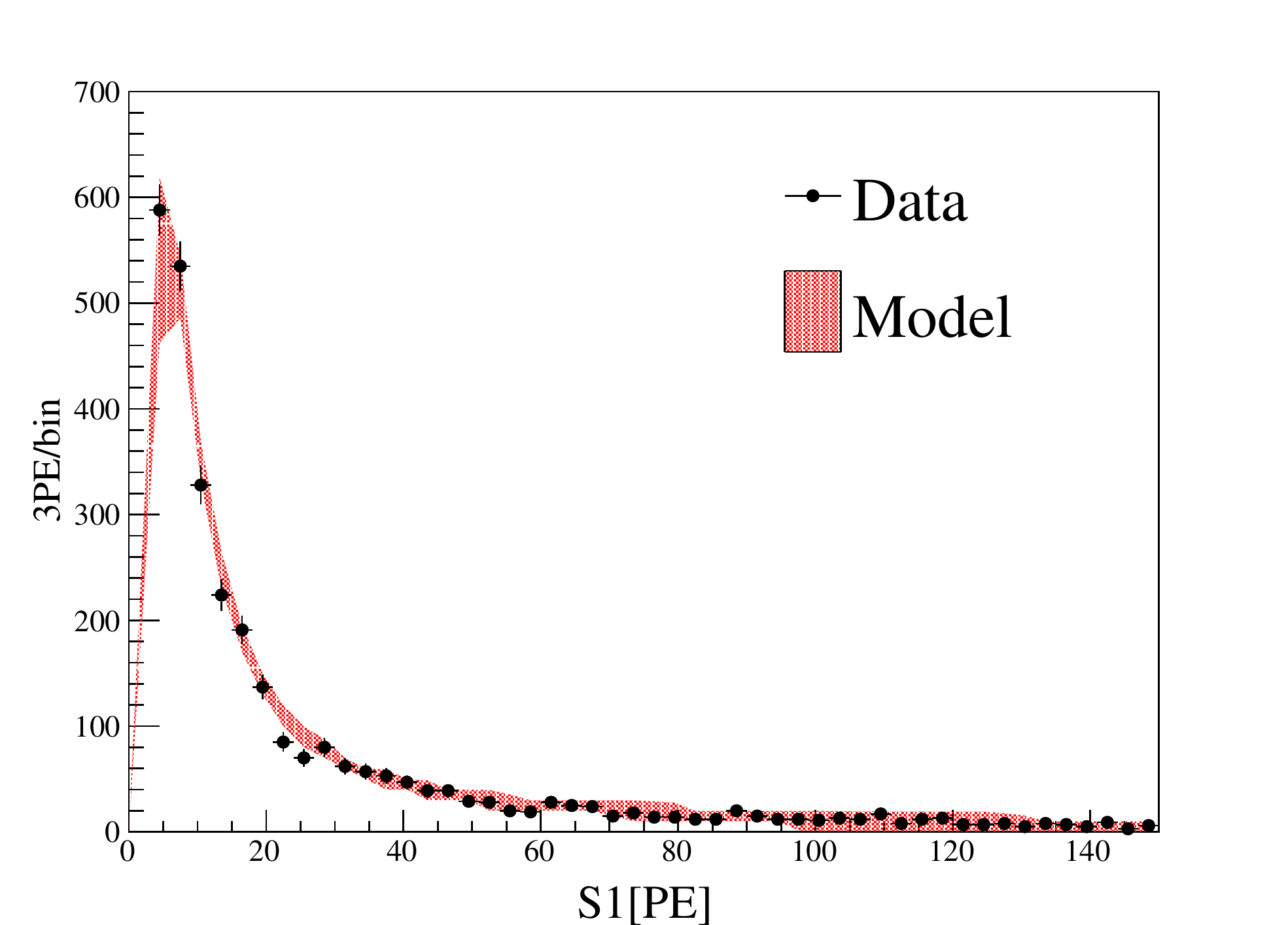}
    \caption{Run 9 AmBe $S1$}
  \end{subfigure}
  \begin{subfigure}{0.3\textwidth}
    \includegraphics[width=1\textwidth]{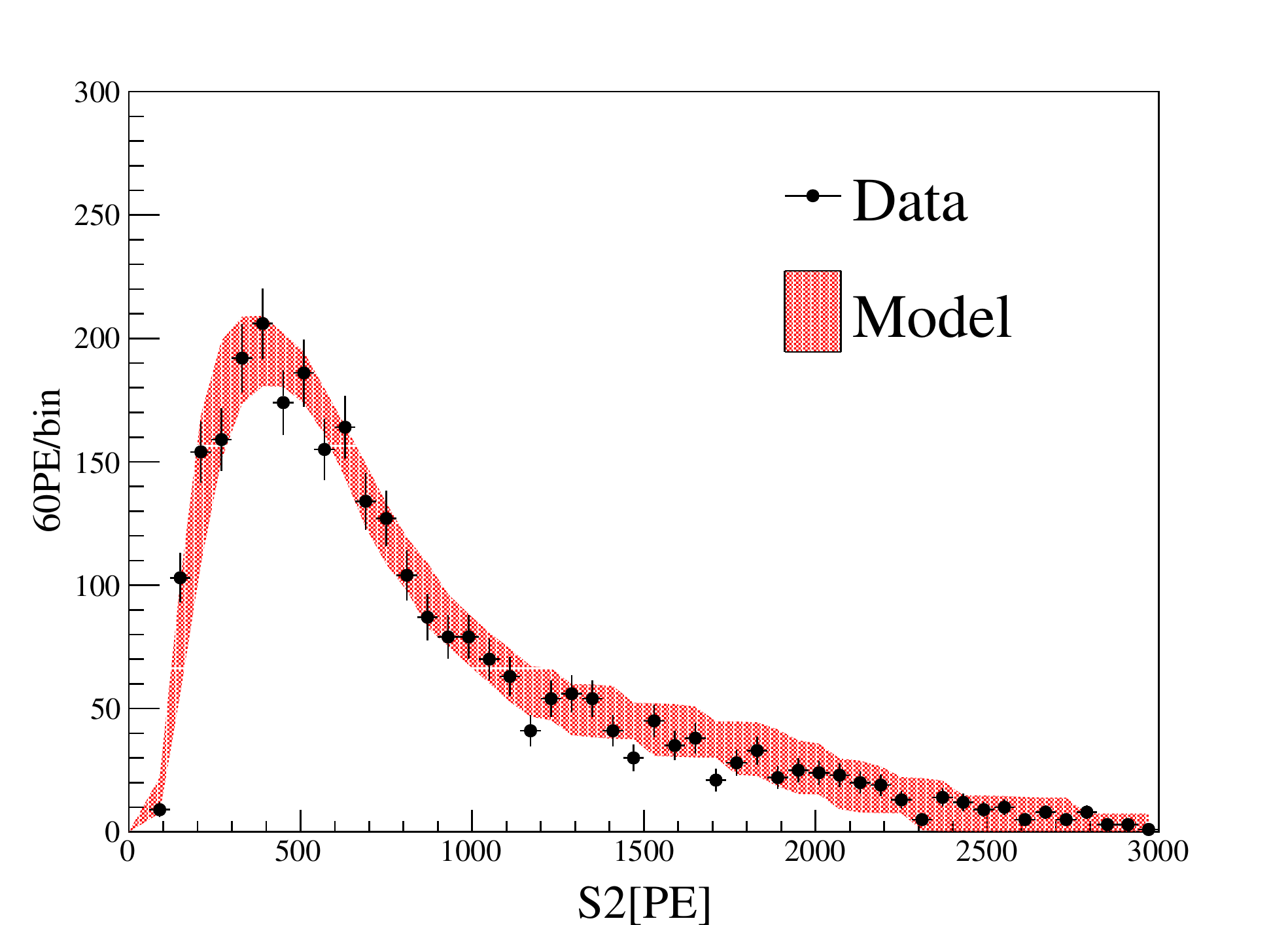}
    \caption{Run 9 AmBe $S2$}
  \end{subfigure}
  \begin{subfigure}{0.3\textwidth}
    \includegraphics[width=1\textwidth]{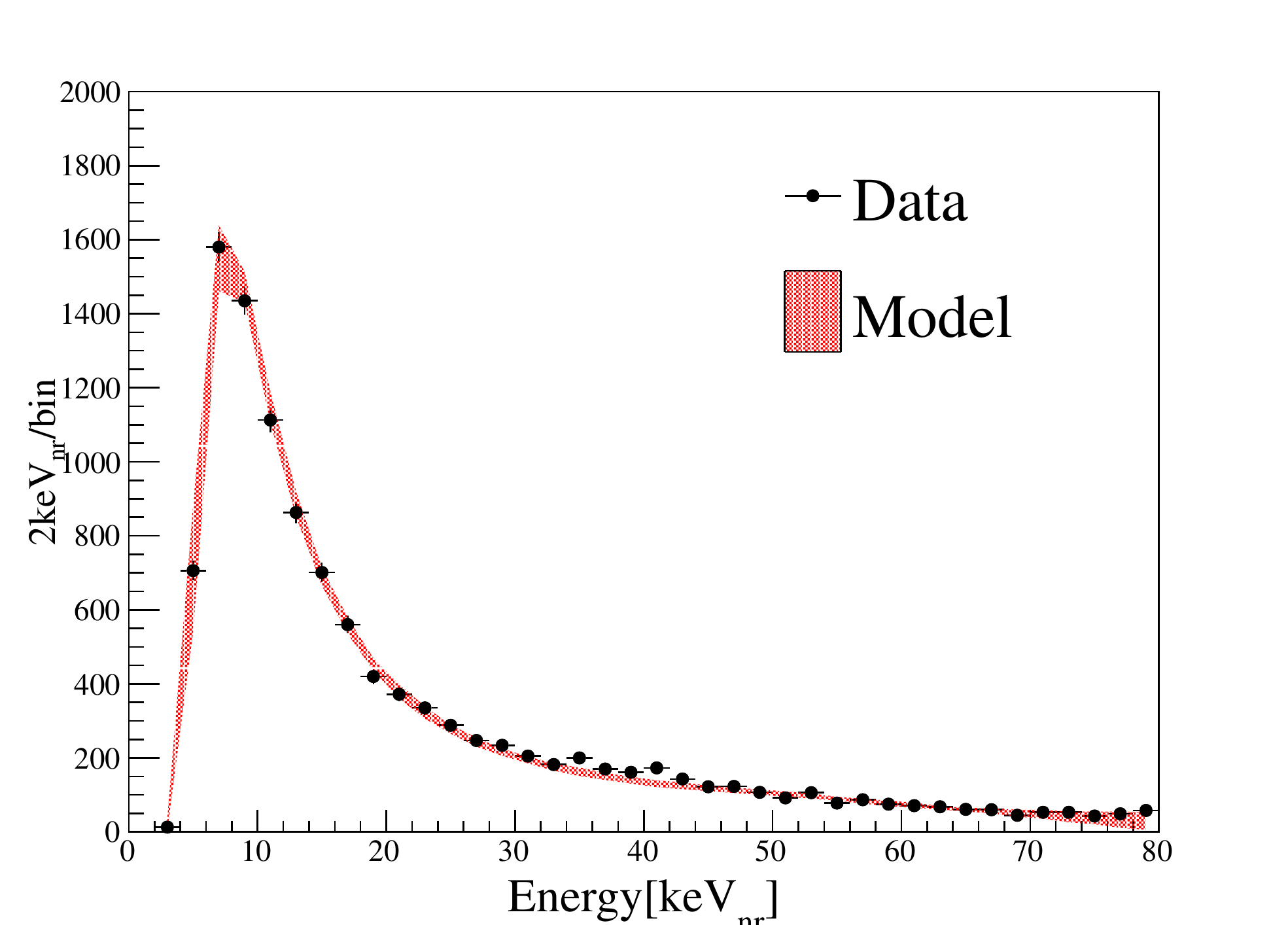}
    \caption{Runs 10/11 AmBe energy}
  \end{subfigure}
  \begin{subfigure}{0.3\textwidth}
    \includegraphics[width=1\textwidth]{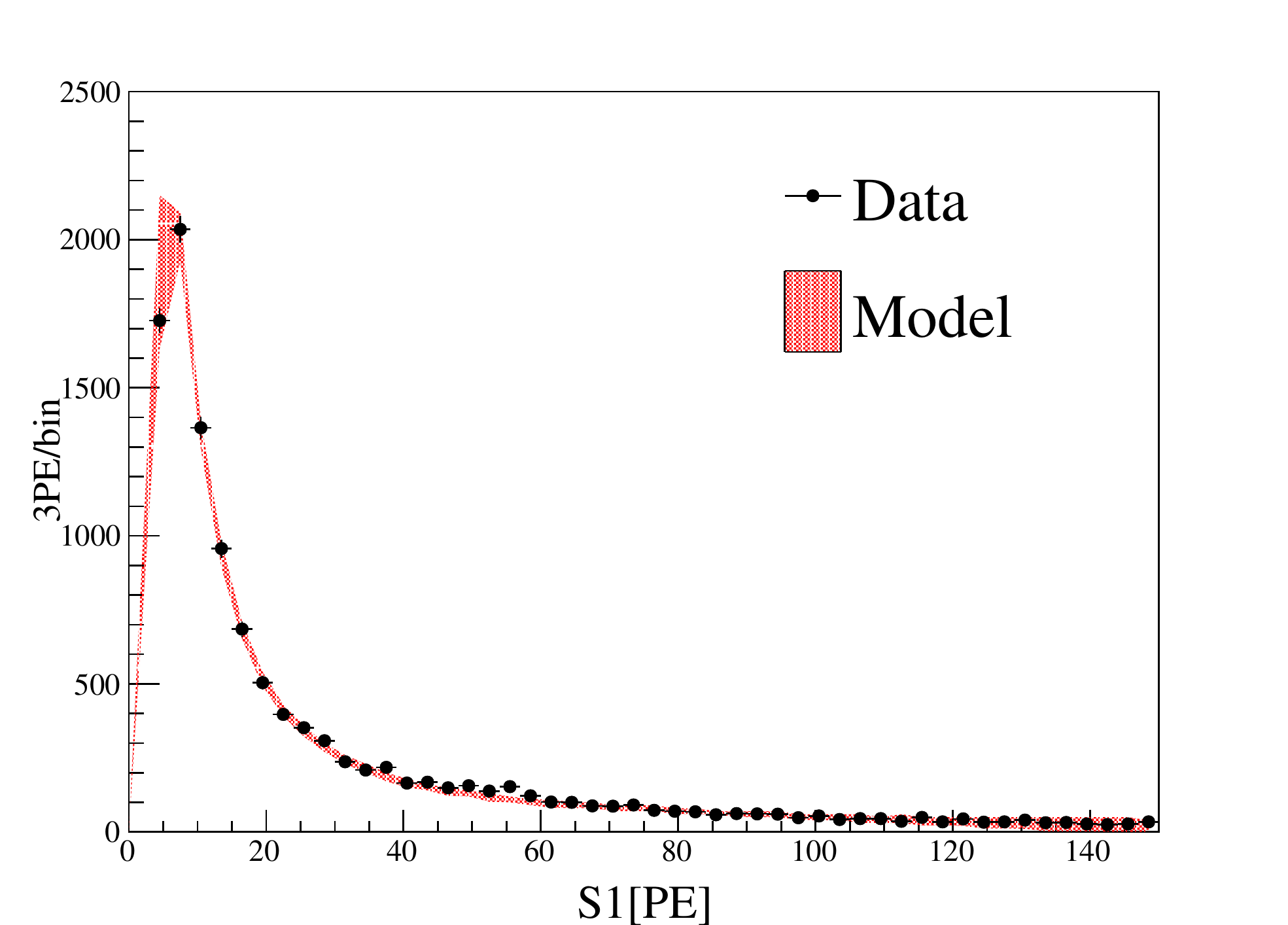}
    \caption{Runs 10/11 AmBe $S1$}
  \end{subfigure}
  \begin{subfigure}{0.3\textwidth}
    \includegraphics[width=1\textwidth]{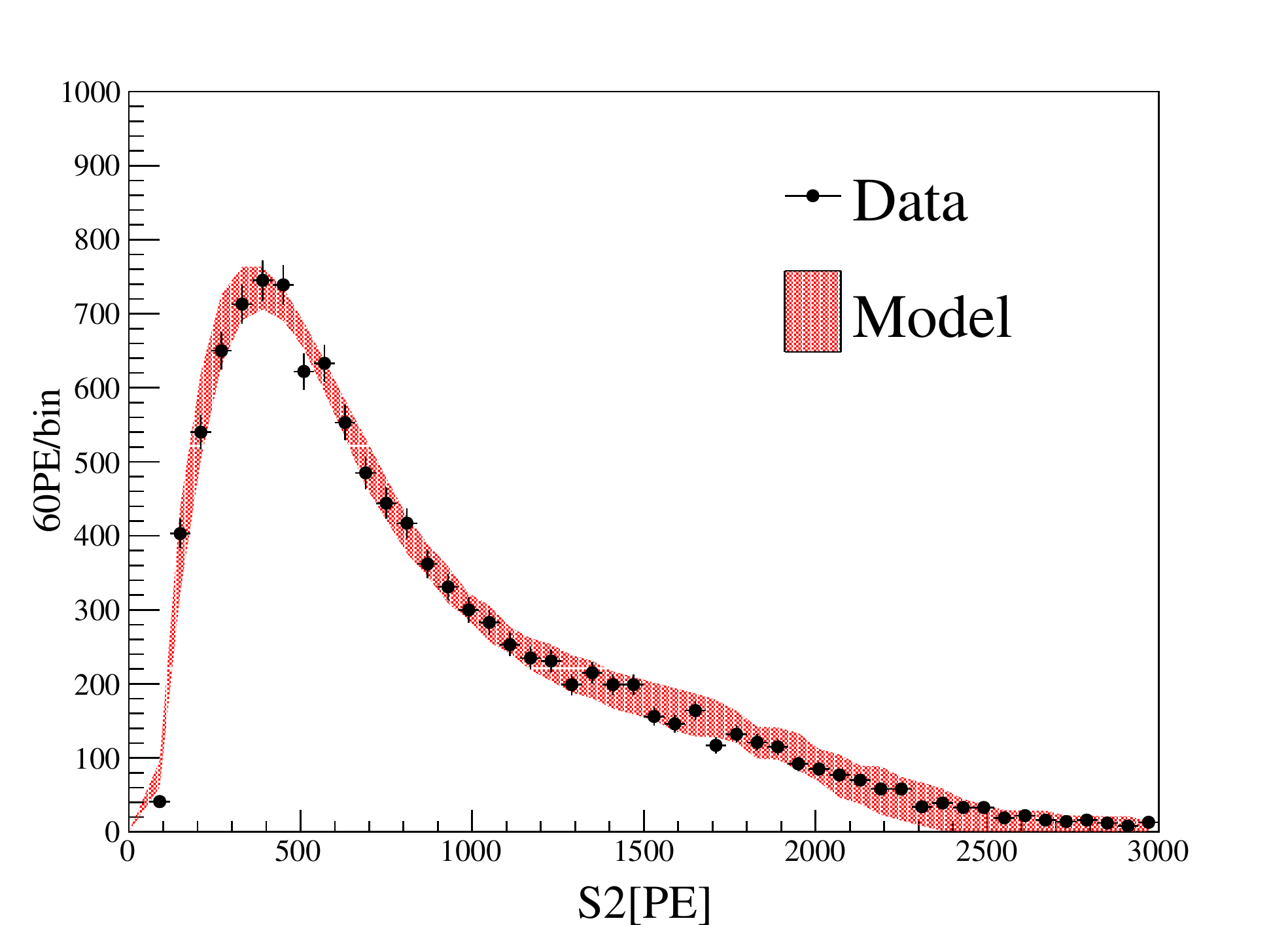}
    \caption{Runs 10/11 AmBe $S2$}
  \end{subfigure}
  \begin{subfigure}{0.3\textwidth}
    \includegraphics[width=1\textwidth]{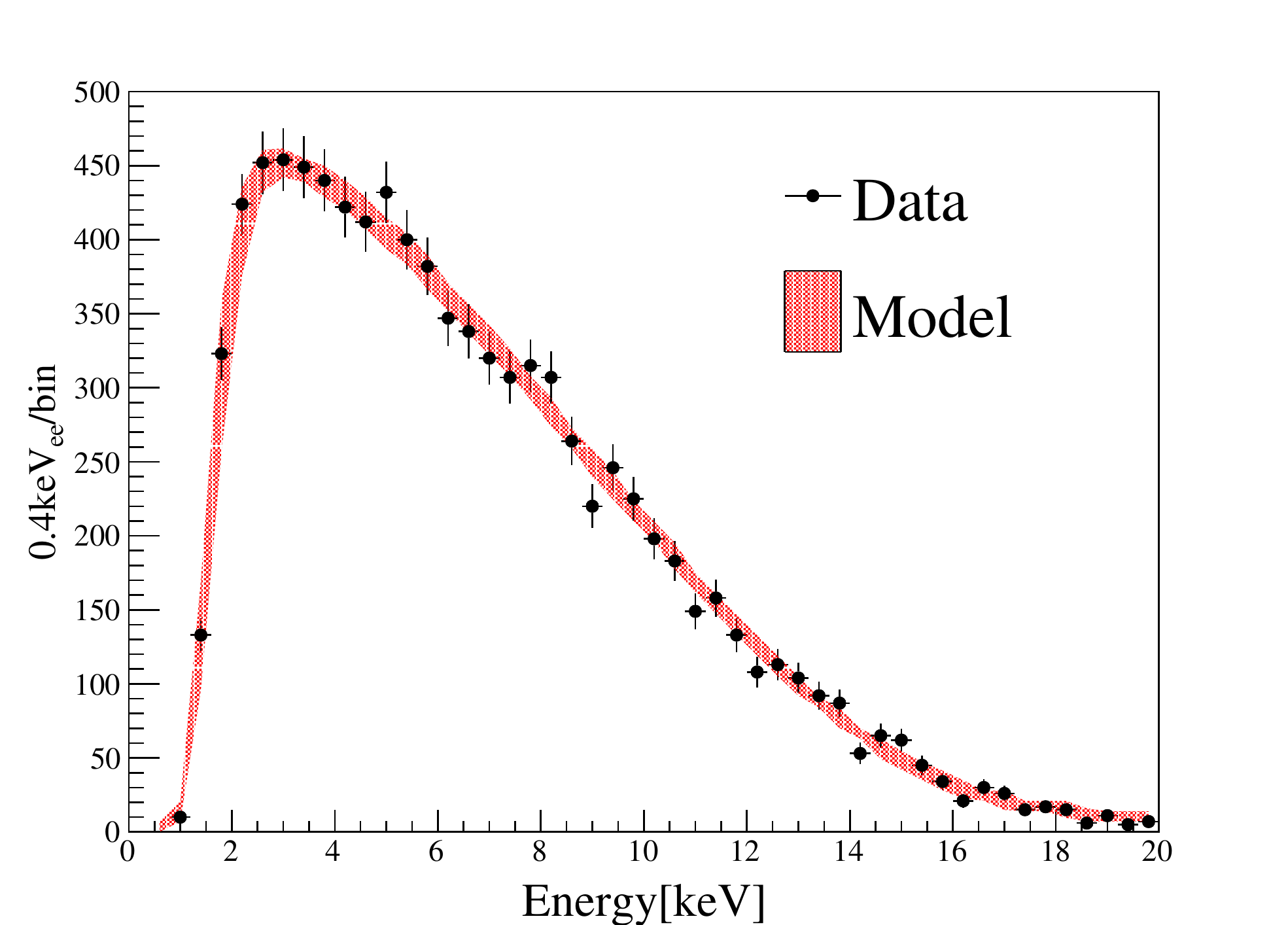}
    \caption{Run 9 CH$^{3}$T energy}
  \end{subfigure}  
  \begin{subfigure}{0.3\textwidth}
    \includegraphics[width=1\textwidth]{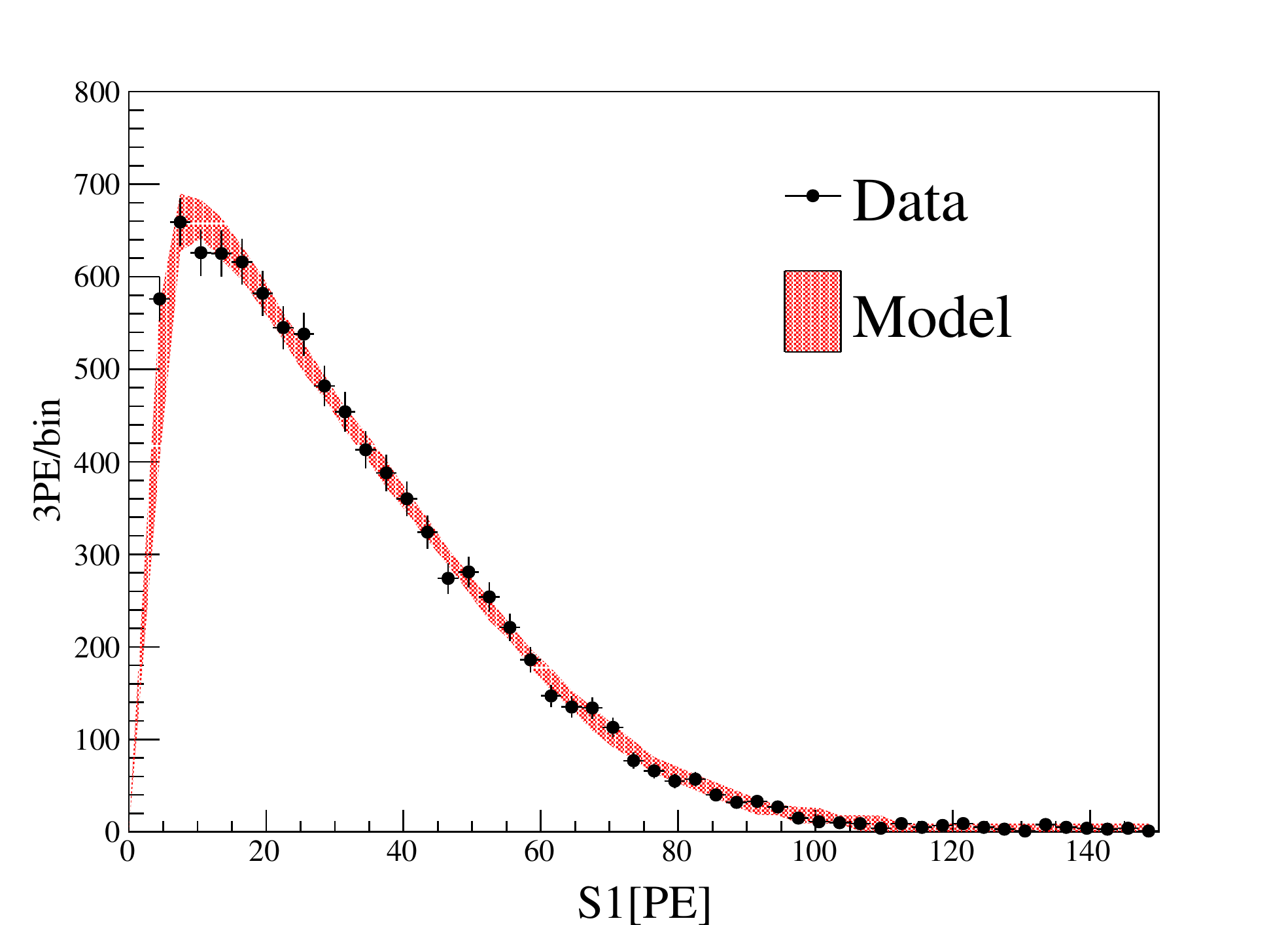}
    \caption{Run 9 CH$^{3}$T $S1$}
  \end{subfigure}  
  \begin{subfigure}{0.3\textwidth}
    \includegraphics[width=1\textwidth]{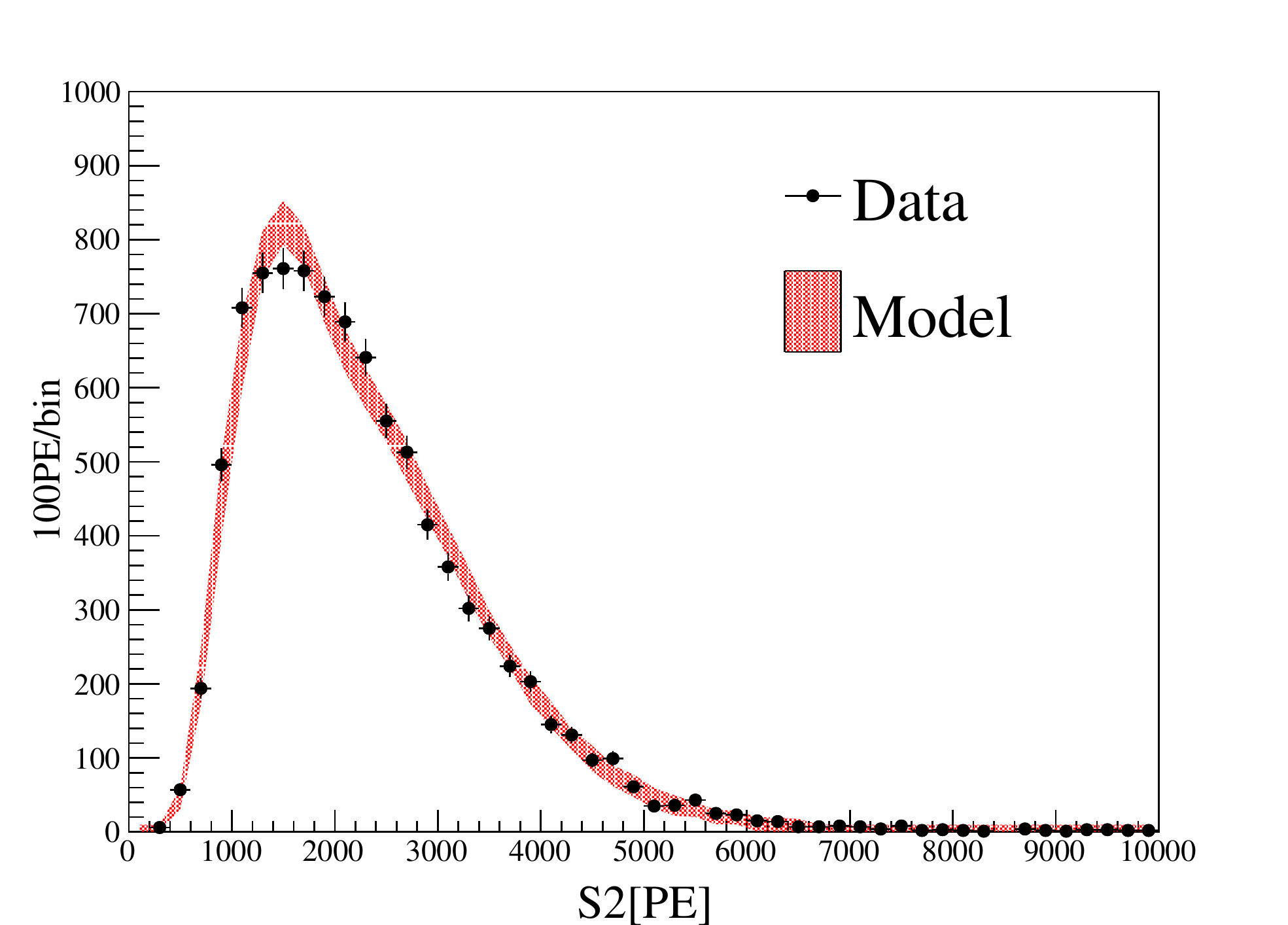}
    \caption{Run 9 CH$^{3}$T $S2$}
  \end{subfigure}  
  \begin{subfigure}{0.3\textwidth}
    \includegraphics[width=1\textwidth]{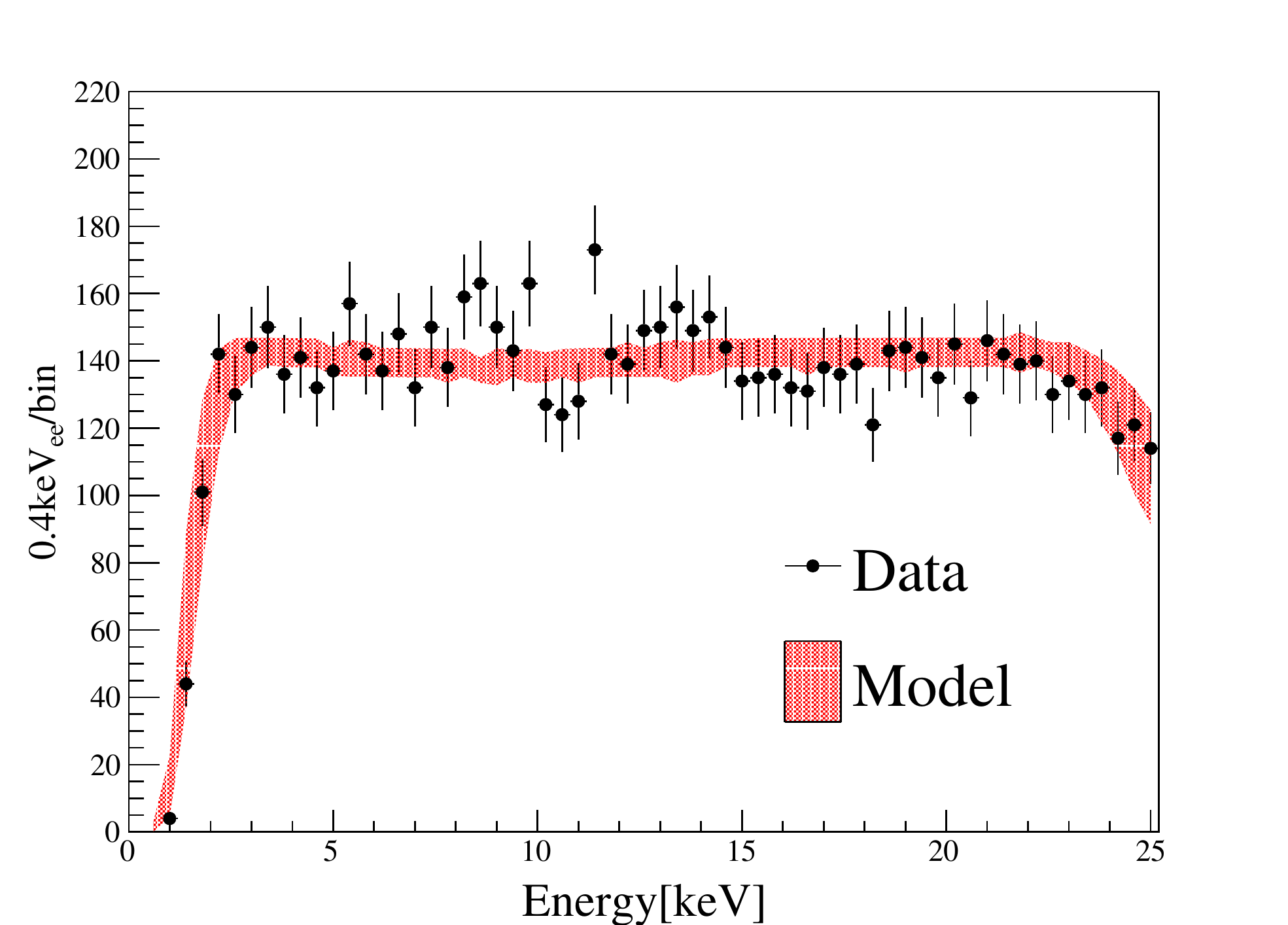}
    \caption{Runs 10/11 $^{220}$Rn energy}
  \end{subfigure}  
  \begin{subfigure}{0.3\textwidth}
    \includegraphics[width=1\textwidth]{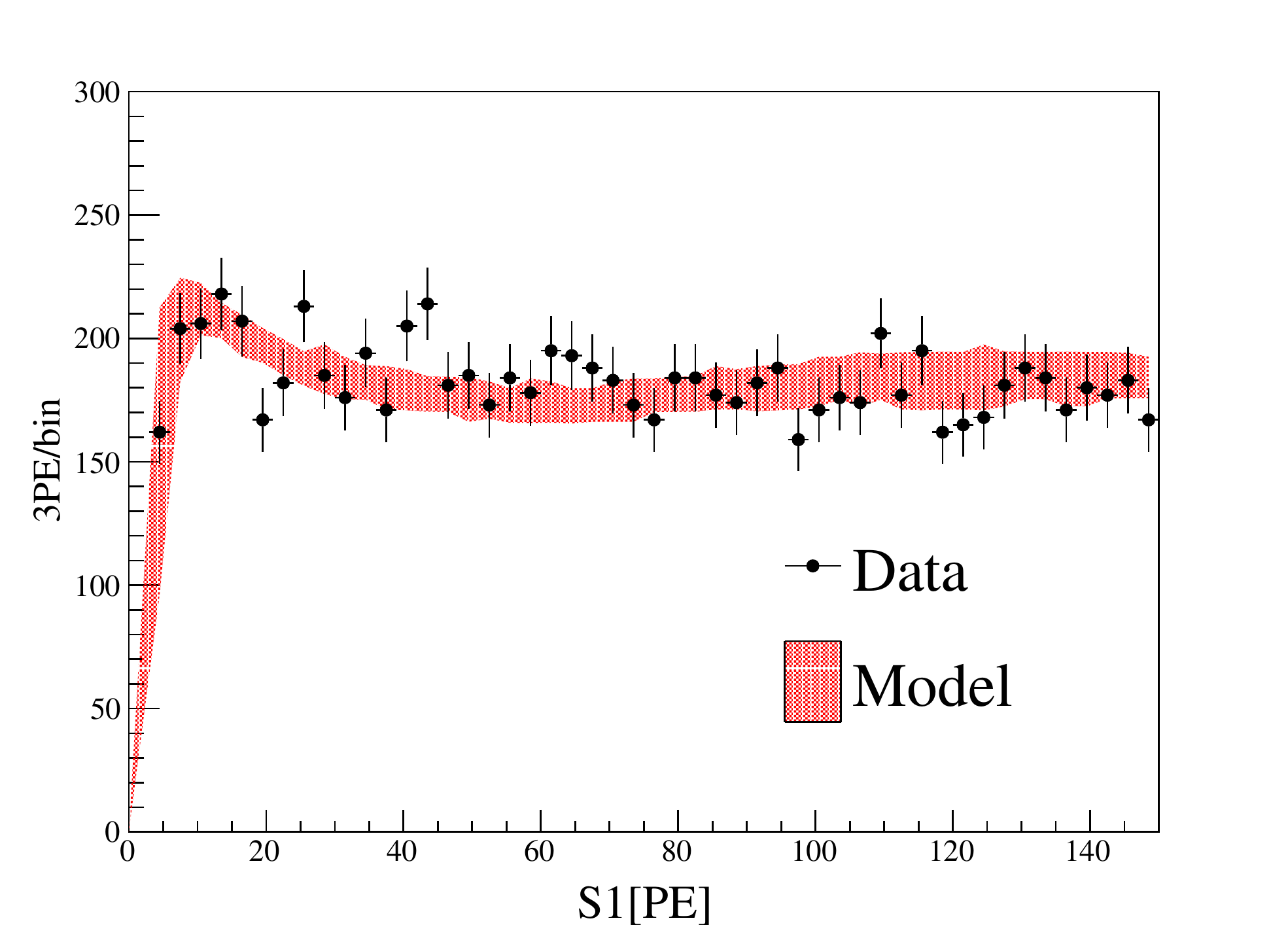}
    \caption{Runs 10/11 $^{220}$Rn $S1$}
  \end{subfigure}  
  \begin{subfigure}{0.3\textwidth}
    \includegraphics[width=1\textwidth]{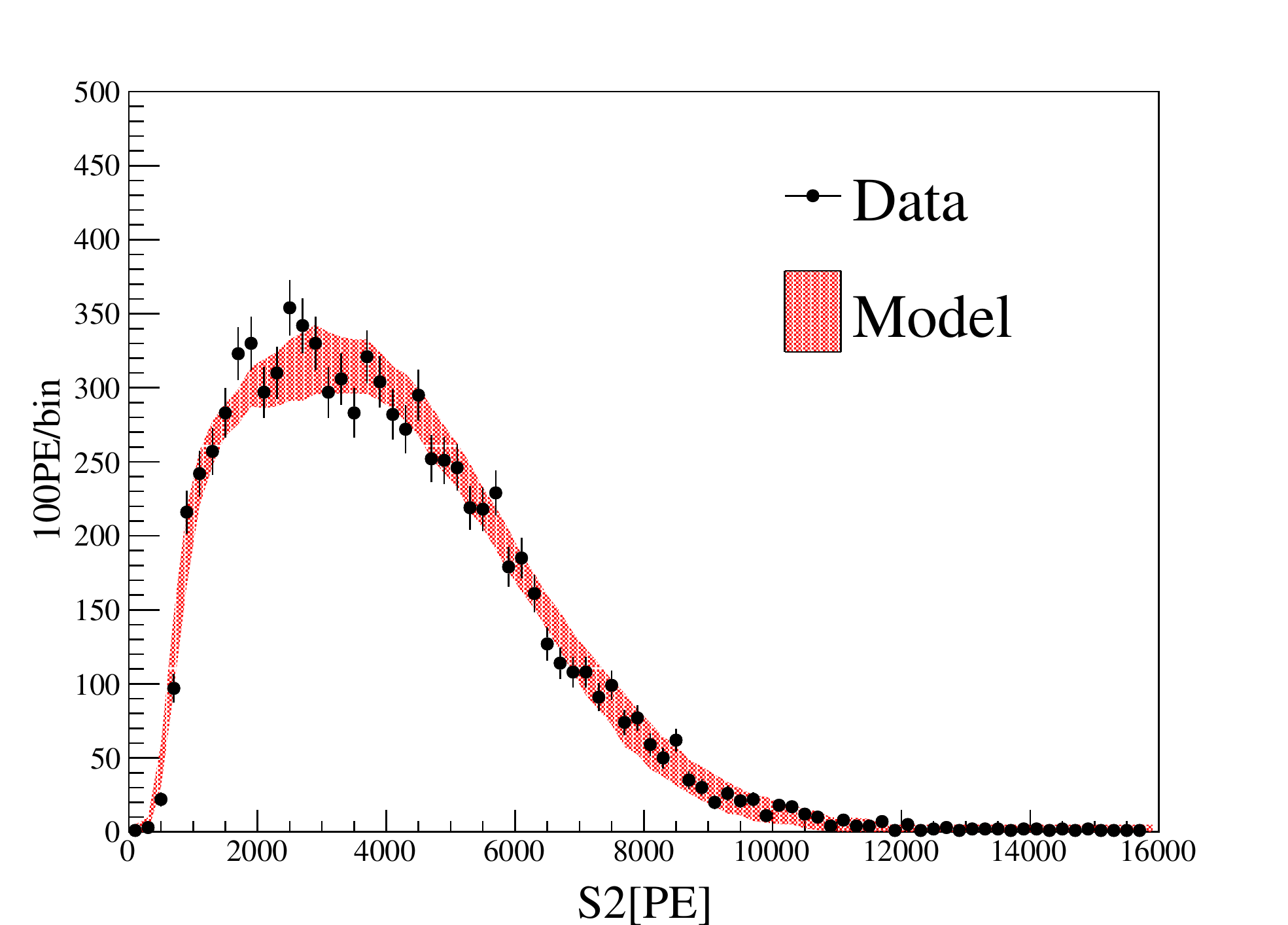}
    \caption{Runs 10/11 $^{220}$Rn $S2$}
  \end{subfigure}  
  \caption{The comparison of calibration data (points) and model (shaded bands = 90\% allowable) in the recoil energy, $S1$, and $S2$, 
  in Run 9 and Runs 10/11. For the $^{220}$Rn energy distribution (j), the decrease at high energy end is due to the 150~PE $S1$ range cut.}
  \label{fig:ER_NR_comparison}
\end{figure}

The parameter space allowable by the calibration data is determined based on the likelihood ratio approach in Ref.~\cite{Cowan:2010js}.
For each set of fixed parameters, 1000 mock data runs are produced with equal but Poisson fluctuated statistics 
as the calibration data. The test statistic for each mock run is defined as the 
difference between the log likelihood calculated using this fixed point PDF, and the global minimum value from the parameter scan,
\begin{equation}
\Delta \mathcal{L} = -2{\ln} \mathcal{L}_{\rm fixed} - (-2{\ln}\mathcal{L}_{\rm min})\,.
\end{equation}
The distributions of $\Delta \mathcal{L}$ for the mock data generated from the best fit parameters for the four calibration data sets are shown in Fig.~\ref{fig:acceptance region}.
The blue dashed regions refer to the 90$\%$ integrals from zero, beyond which 
the difference between the mock data set and its own PDF becomes less likely. It is verified that
the 90\% boundary values for $\Delta \mathcal{L}$ at other parameter space points are similar.
Therefore, $\Delta \mathcal{L}$ of the real data is tested around the best fit, and the allowable space is defined by the 90$\%$ boundaries in Fig.~\ref{fig:acceptance region}.
The corresponding allowable range of distributions in recoil energy, $S1$, and $S2$ are shown in 
Fig.~\ref{fig:ER_NR_comparison}, together with the calibration data, where good agreements are found.

% calibration data part
%By comparison calibration data and simulation, the best fit model is found corresponding to the global minimum $\mathcal{L}_0$.
%The best fit $g_1$ and $g_2$ are 0.1135 and 11.29 for Run 9 and 0.1199 and 11.11 for Runs 10/11 
%within our nominal values in Table~\ref{tab:datasets}. The best efficiency for $S1$ is 
%found at $\epsilon_1 =1/\left({1+\exp(\frac{S1-3.1}{0.075})}\right)$ (Run 9)
%and $\epsilon_1 = 1/(\left({1+\exp(\frac{S1-4.0}{0.8})}\right)$(Runs 
%10/11), adopted already in Ref.~\cite{Wang:2020coa}.
%The ones with likelihood of within $\mathcal{L}_0$ + $\Delta \mathcal{L}$ is accepted. 
%The comparison of data and simulation is shown in Fig.~\ref{fig:ER_NR_comparison}, the data is in good agreement with simulation.

%application in dark matter search

\begin{figure}[!htbp]
  \centering
  \begin{subfigure}{0.49\textwidth}
    \includegraphics[width=1.0\textwidth]{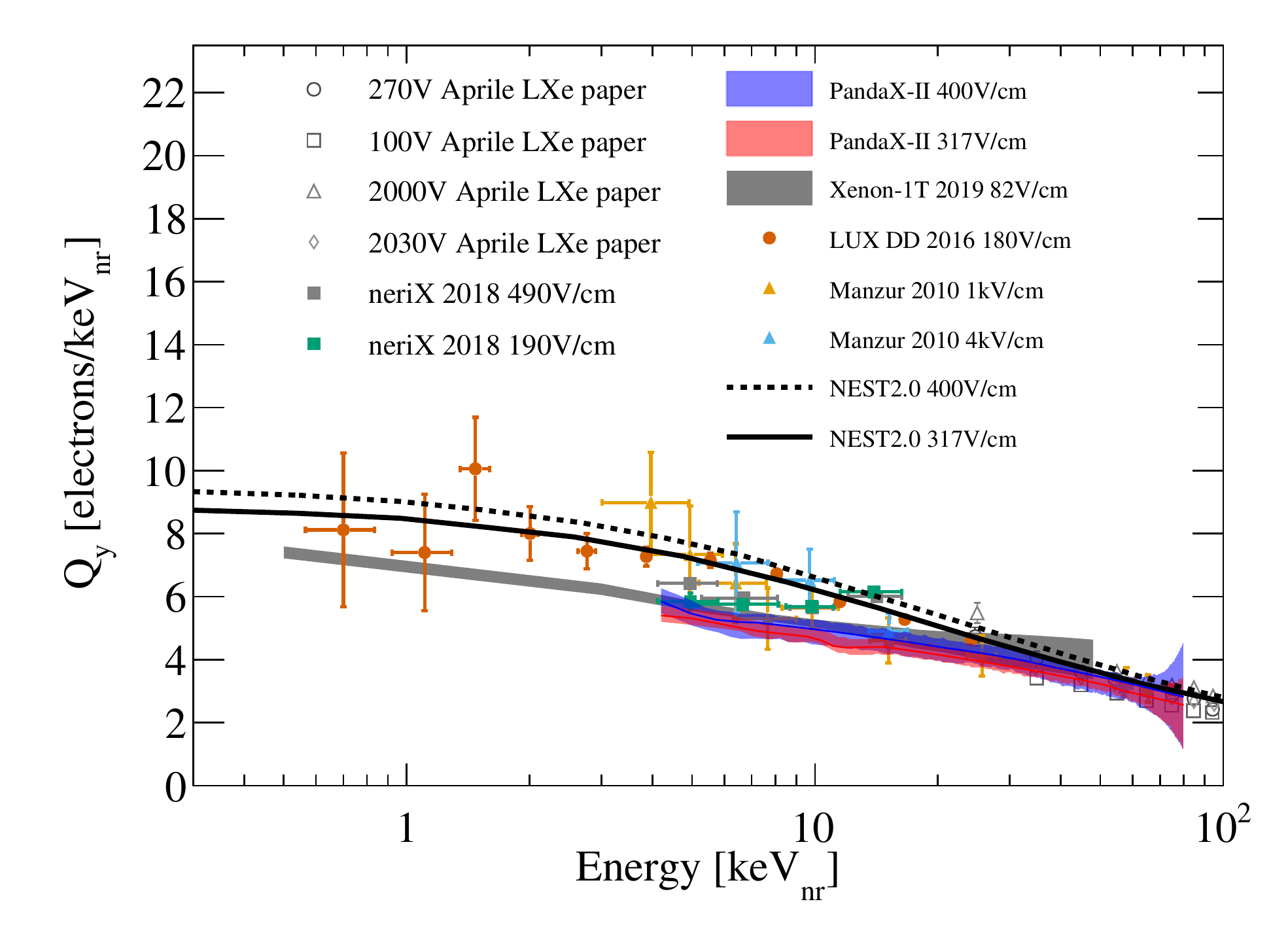}
    \caption{$Q_y$ of NR}
    \label{fig:cy_nr_worlddata}
  \end{subfigure}
    \begin{subfigure}{0.49\textwidth}
    \includegraphics[width=1.0\textwidth]{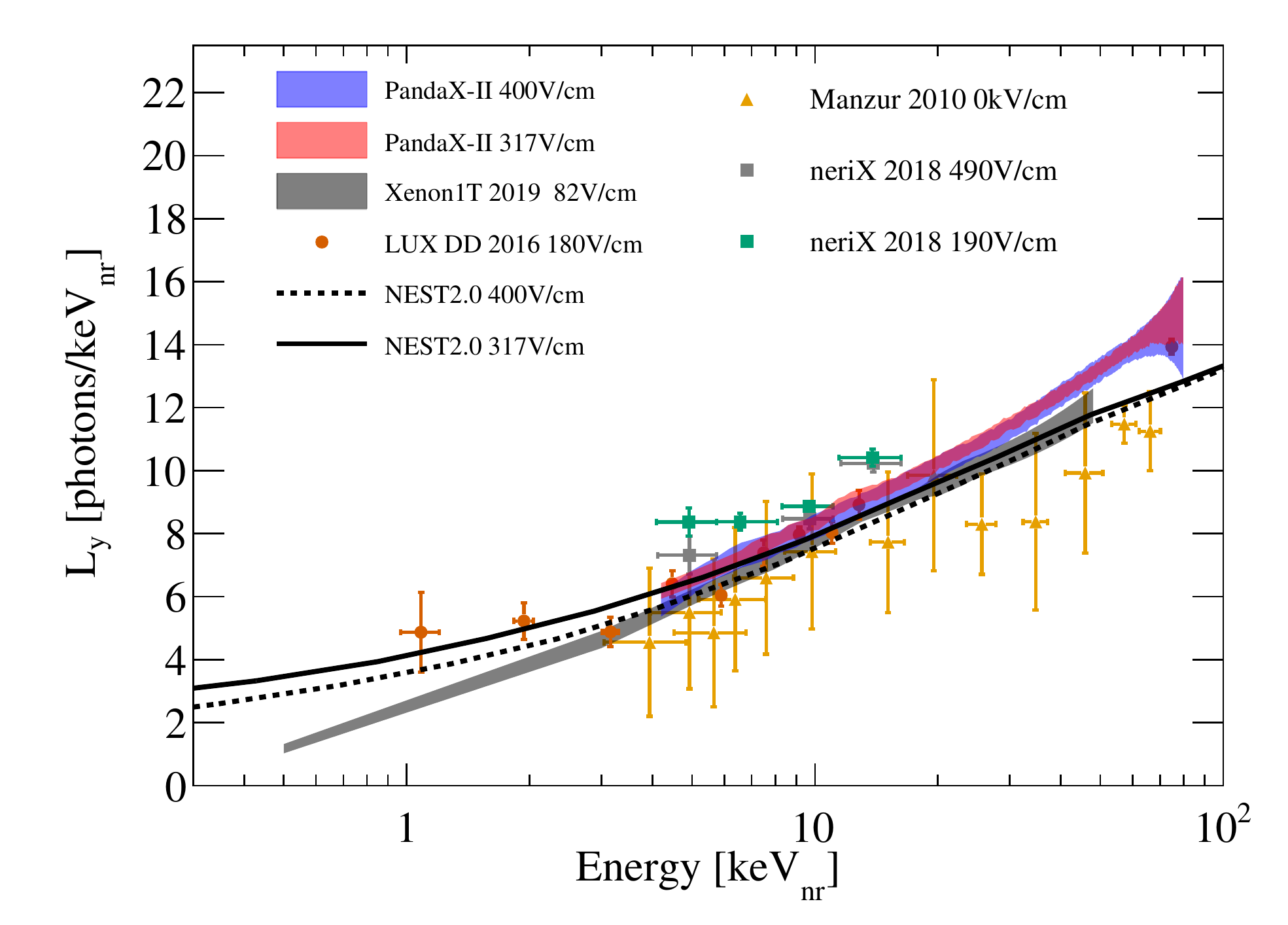}
    \caption{$L_y$ of NR}
    \label{fig:ly_nr_worlddata}
  \end{subfigure}
  \begin{subfigure}{0.49\textwidth}
    \includegraphics[width=1.0\textwidth]{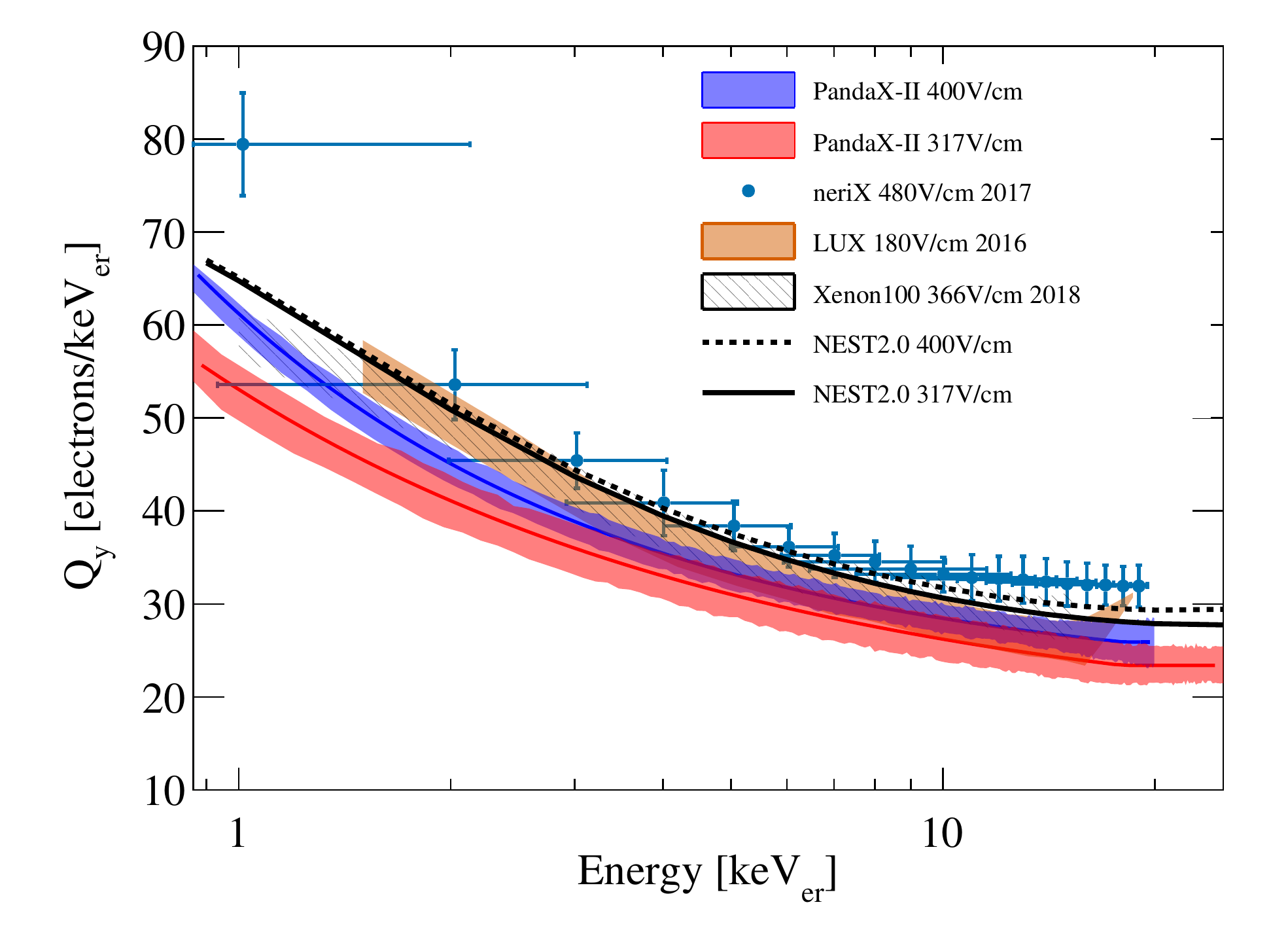}
    \caption{$Q_y$ of ER}
    \label{fig:cy_er_worlddata}
  \end{subfigure}
  \begin{subfigure}{0.49\textwidth}
    \includegraphics[width=1.0\textwidth]{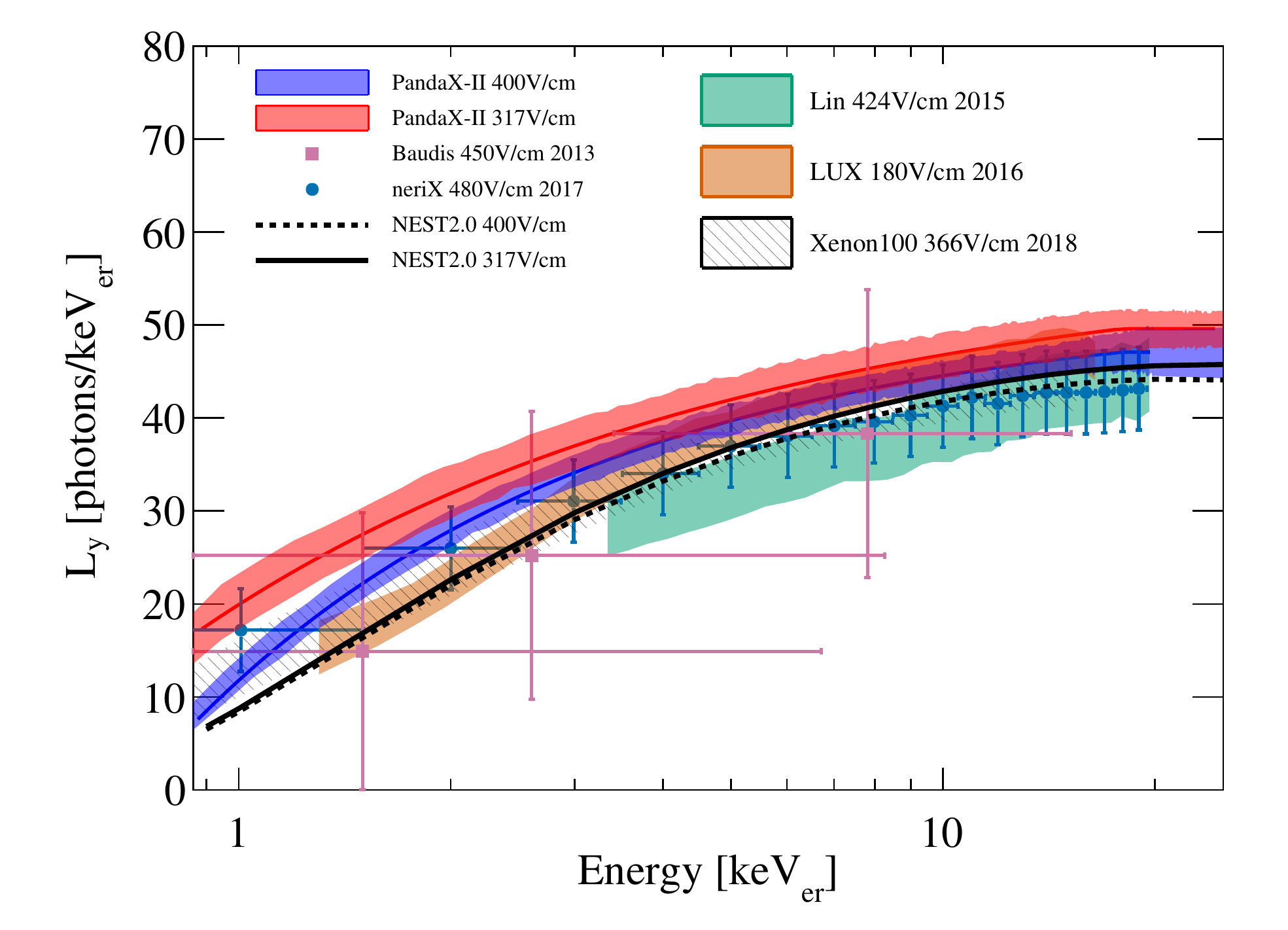}
    \caption{$L_y$ of ER}
    \label{fig:ly_er_worlddata}
  \end{subfigure}
  \caption{Charge yield $Q_y$ (a) and light yield $L_y$ (b) of the NR and $Q_y$ (c) and $L_y$ (d) of the ER obtained from
    the PandaX-II data: blue=400 V/cm, red=317 V/cm. Overlaid world data include: NR from Refs.~\cite{Aprile:2006kx,Manzur:2009hp,Sorensen:2010hq,Aprile:2013teh,Akerib:2016mzi,Aprile:2018jvg},
    and ER from Refs.~\cite{Baudis:2013cca,Lin:2015jta,Akerib:2015wdi,Goetzke:2016lfg}, as indicated in the legend. The native NEST2.0 predictions are drawn in black curves, solid (317 V/cm), and dashed (400 V/cm).
    The XENON1T responses~\cite{Aprile:2019dme} 
    are not included in the ER figures since the operation field (81 V/cm) is significantly different from the PandaX-II conditions, and for visual clarity.
}
  \label{fig:yield curves world data}
\end{figure}

The resulting best fit $Q_y$ and $L_y$ for the NR and ER events are shown in Fig.~\ref{fig:yield curves world data}, 
overlaid with the world data, as well as the native NEST2.0 predictions~\cite{NEST2p0}.
%Since $L_y$ and $Q_y$ are correlated, the plot choices follow conventions adopted by the majority of the world data. 
The shaded bands
indicate the 90\% allowable model space, with uncertainties due to detector parameters and 
statistics of the calibration data naturally incorporated. Our NR models cover a wide energy 
range from 4 to 80 keV$_{\rm nr}$.
At the two drift fields (400 V/cm and 317 V/cm), our best NR models are consistent as expected.
For the $Q_y$ distribution with recoil energy from 4 to 15 keV$_{\rm nr}$, there is 
significant spread among the world data, in which our $Q_y$
appears to be in better agreement with Ref.~\cite{Aprile:2019dme} (Xenon-1T 2019), 
but lower than the others. The 
NEST2.0 global fit, presumably mostly driven by data from Ref.~\cite{Akerib:2016mzi} (LUX DD), 
has a higher $Q_y$ than ours.
The global data agreement improves significantly above 15 keV$_{\rm nr}$.
$L_y$ of our NR models, on the other hand, appear to be in agreement with most of the world data, except
some slight tension at above 25 keV$_{\rm nr}$ with Ref.~\cite{Manzur:2009hp} (Manzur 2010), which bears large uncertainties by itself. 

For the ER models, $Q_y$ ($L_y$) for Run 9 is higher (lower) than that for Runs 10/11. Such a behavior can also be expected since the initial ionized electrons are less likely to be recombined in stronger drift field.
Our model at 400~V/cm is in reasonable agreement with Ref.~\cite{Aprile:2017xxh} (Xenon100)
at similar drift field, but is in some tension with Refs.~\cite{Goetzke:2016lfg, Lin:2015jta} (neriX~480~V/cm, Lin~424~V/cm).
Our $Q_y$ ($L_y$) at 317 V/cm is generally lower (higher) than the world data, including that from 
Ref.~\cite{Akerib:2016mzi} (LUX) taken at 180~V/cm (and that from Ref.~\cite{Aprile:2019dme}
at 81~V/cm, not drawn), as well as the native NEST2.0 predictions. 
Given the uncertainties in all these measurement, however, more systematic studies and comparisons are warranted.

Regardless of the global comparison, it should be emphasized that for PandaX-II, models determined from {\it in situ} 
calibrations are the most self-consistent models to be used in the dark matter search data. 
Our best fit models presented here have therefore been adopted in the analysis in Ref.~\cite{Wang:2020coa}. 

\section{Conclusion}
\label{sec:conclusion}
We report the ER and NR responses from the PandaX-II detector based on 
all calibration data obtained during the operation at 
two different drift fields (400 V/cm and 317 V/cm). The empirical best fits to the data and model 
uncertainties are obtained, yielding good agreements between the data and our models.
In comparison to those presented in Refs.~\cite{Tan:2016zwf,Cui:2017nnn}, the models in this work
cover the entire PandaX-II data taking period, with a more extend energy range between 4 to 80~keV$_{\rm nr}$(NR) and 
1 to 25~keV$_{\rm ee}$(ER). At the two drift fields, our NR models are 
in agreement, and our ER models exhibit a relative shift. Both behaviors are consistent with expectation.

Our models are also compared to the world data. Our NR models lie within the large global spread. 
For the ER response, our model yields a higher (lower) $L_y$ ($Q_y$) in comparison to most of the world data, 
indicating some unaccounted systematic uncertainties in our or others' measurements. These discrepancies 
encourage continuous calibration effort and further investigations of systematics 
in the data. Finally, the analysis approach presented here is general 
and can be applied to similar noble liquid TPC experiments.

\section{Acknowledgement}
This project is supported in part by a grant from the Ministry of Science and Technology of
  China (No. 2016YFA0400301), grants from National Science
  Foundation of China (Nos. 12090060, 11525522, 11775141 and
  11755001), and Office of Science and
  Technology, Shanghai Municipal Government (grant No. 18JC1410200). We thank supports from Double First Class Plan of
  the Shanghai Jiao Tong University. We also thank the sponsorship from the
  Chinese Academy of Sciences Center for Excellence in Particle
  Physics (CCEPP), Hongwen Foundation in Hong Kong, and Tencent
  Foundation in China. Finally, we thank the CJPL administration and
  the Yalong River Hydropower Development Company Ltd. for
  indispensable logistical support and other help.

 %\citep{adams1995hitchhiker}

%\bibliographystyle{plain}
\bibliographystyle{unsrt}
\bibliography{main}

\begin{thebibliography}{10}

\bibitem{Tan:2016diz}
Andi Tan et~al.
\newblock {Dark Matter Search Results from the Commissioning Run of PandaX-II}.
\newblock {\em Phys. Rev.}, D93(12):122009, 2016.

\bibitem{Aprile:2017aty}
E.~Aprile et~al.
\newblock {The XENON1T Dark Matter Experiment}.
\newblock {\em Eur. Phys. J. C}, 77(12):881, 2017.

\bibitem{PhysRevLett.112.091303}
D.~S. Akerib et~al.
\newblock First results from the lux dark matter experiment at the sanford
  underground research facility.
\newblock {\em Phys. Rev. Lett.}, 112:091303, Mar 2014.

\bibitem{NEST2p0}
M.~{Szydagis} et~al.
\newblock \url{https://doi.org/10.5281/zenodo.4283077}.

\bibitem{Yu-Cheng:2013iaa}
Yu-Cheng Wu et~al.
\newblock {Measurement of Cosmic Ray Flux in China JinPing underground
  Laboratory}.
\newblock {\em Chin. Phys. C}, 37(8):086001, 2013.

\bibitem{Wang:2020coa}
Qiuhong Wang et~al.
\newblock {Results of Dark Matter Search using the Full PandaX-II Exposure}.
\newblock 7 2020.

\bibitem{Akerib:2015wdi}
D.~S. Akerib et~al.
\newblock {Tritium calibration of the LUX dark matter experiment}.
\newblock {\em Phys. Rev.}, D93(7):072009, 2016.

\bibitem{Aprile:2016pmc}
E.~Aprile et~al.
\newblock {Results from a Calibration of XENON100 Using a Source of Dissolved
  Radon-220}.
\newblock {\em Phys. Rev. D}, 95(7):072008, 2017.

\bibitem{Ma:2020kll}
Wenbo Ma et~al.
\newblock {Internal Calibration of the PandaX-II Detector with Radon Gaseous
  Sources}.
\newblock 6 2020.

\bibitem{Wang:2019opt}
Qiuhong Wang et~al.
\newblock {An Improved Evaluation of the Neutron Background in the PandaX-II
  Experiment}.
\newblock {\em Sci.\ China Phys.\ Mech.\ Astron.}, 63(3):231011, 2020.

\bibitem{Xiao:2015psa}
Xiang Xiao et~al.
\newblock {Low-mass dark matter search results from full exposure of the
  PandaX-I experiment}.
\newblock {\em Phys. Rev.}, D92(5):052004, 2015.

\bibitem{Szydagis:2011tk}
M.~Szydagis, N.~Barry, K.~Kazkaz, J.~Mock, D.~Stolp, M.~Sweany, M.~Tripathi,
  S.~Uvarov, N.~Walsh, and M.~Woods.
\newblock {NEST: A Comprehensive Model for Scintillation Yield in Liquid
  Xenon}.
\newblock {\em JINST}, 6:P10002, 2011.

\bibitem{NEST1.0NR}
B.~{Lenardo}, K.~{Kazkaz}, A.~{Manalaysay}, J.~{Mock}, M.~{Szydagis}, and
  M.~{Tripathi}.
\newblock A global analysis of light and charge yields in liquid xenon.
\newblock {\em IEEE Transactions on Nuclear Science}, 62(6):3387--3396, 2015.

\bibitem{Li:2015qhq}
Shaoli Li et~al.
\newblock {Performance of Photosensors in the PandaX-I Experiment}.
\newblock {\em JINST}, 11(02):T02005, 2016.

\bibitem{Wu:2017cjl}
Qinyu Wu et~al.
\newblock {Update of the trigger system of the PandaX-II experiment}.
\newblock {\em JINST}, 12(08):T08004, 2017.

\bibitem{Cui:2017nnn}
Xiangyi Cui et~al.
\newblock {Dark Matter Results From 54-Ton-Day Exposure of PandaX-II
  Experiment}.
\newblock {\em Phys. Rev. Lett.}, 119(18):181302, 2017.

\bibitem{Cowan:2010js}
Glen Cowan, Kyle Cranmer, Eilam Gross, and Ofer Vitells.
\newblock {Asymptotic formulae for likelihood-based tests of new physics}.
\newblock {\em Eur. Phys. J. C}, 71:1554, 2011.
\newblock [Erratum: Eur.Phys.J.C 73, 2501 (2013)].

\bibitem{Aprile:2019dme}
E.~Aprile et~al.
\newblock {XENON1T dark matter data analysis: Signal and background models and
  statistical inference}.
\newblock {\em Phys. Rev. D}, 99(11):112009, 2019.

\bibitem{Akerib:2016mzi}
D.~S. Akerib et~al.
\newblock {Low-energy (0.7-74 keV) nuclear recoil calibration of the LUX dark
  matter experiment using D-D neutron scattering kinematics}.
\newblock 2016.

\bibitem{Manzur:2009hp}
A.~Manzur, A.~Curioni, L.~Kastens, D.N. McKinsey, K.~Ni, and T.~Wongjirad.
\newblock {Scintillation efficiency and ionization yield of liquid xenon for
  mono-energetic nuclear recoils down to 4 keV}.
\newblock {\em Phys. Rev. C}, 81:025808, 2010.

\bibitem{Aprile:2017xxh}
E.~Aprile et~al.
\newblock {Signal Yields of keV Electronic Recoils and Their Discrimination
  from Nuclear Recoils in Liquid Xenon}.
\newblock {\em Phys. Rev. D}, 97(9):092007, 2018.

\bibitem{Goetzke:2016lfg}
L.W. Goetzke, E.~Aprile, M.~Anthony, G.~Plante, and M.~Weber.
\newblock {Measurement of light and charge yield of low-energy electronic
  recoils in liquid xenon}.
\newblock {\em Phys. Rev. D}, 96(10):103007, 2017.

\bibitem{Lin:2015jta}
Qing Lin, Jialing Fei, Fei Gao, Jie Hu, Yuehuan Wei, Xiang Xiao, Hongwei Wang,
  and Kaixuan Ni.
\newblock {Scintillation and ionization responses of liquid xenon to low energy
  electronic and nuclear recoils at drift fields from 236 V/cm to 3.93 kV/cm}.
\newblock {\em Phys. Rev. D}, 92(3):032005, 2015.

\bibitem{Aprile:2006kx}
E.~Aprile, C.E. Dahl, L.~DeViveiros, R.~Gaitskell, K.L. Giboni, J.~Kwong,
  P.~Majewski, Kaixuan Ni, T.~Shutt, and M.~Yamashita.
\newblock {Simultaneous measurement of ionization and scintillation from
  nuclear recoils in liquid xenon as target for a dark matter experiment}.
\newblock {\em Phys. Rev. Lett.}, 97:081302, 2006.

\bibitem{Sorensen:2010hq}
Peter Sorensen.
\newblock {A coherent understanding of low-energy nuclear recoils in liquid
  xenon}.
\newblock {\em JCAP}, 09:033, 2010.

\bibitem{Aprile:2013teh}
E.~Aprile et~al.
\newblock {Response of the XENON100 Dark Matter Detector to Nuclear Recoils}.
\newblock {\em Phys. Rev. D}, 88:012006, 2013.

\bibitem{Aprile:2018jvg}
E.~Aprile, M.~Anthony, Q.~Lin, Z.~Greene, P.~De~Perio, F.~Gao, J.~Howlett,
  G.~Plante, Y.~Zhang, and T.~Zhu.
\newblock {Simultaneous measurement of the light and charge response of liquid
  xenon to low-energy nuclear recoils at multiple electric fields}.
\newblock {\em Phys. Rev. D}, 98(11):112003, 2018.

\bibitem{Baudis:2013cca}
Laura Baudis, Hrvoje Dujmovic, Christopher Geis, Andreas James, Alexander Kish,
  Aaron Manalaysay, Teresa Marrodan~Undagoitia, and Marc Schumann.
\newblock {Response of liquid xenon to Compton electrons down to 1.5 keV}.
\newblock {\em Phys. Rev. D}, 87(11):115015, 2013.

\bibitem{Tan:2016zwf}
Andi Tan et~al.
\newblock {Dark Matter Results from First 98.7 Days of Data from the PandaX-II
  Experiment}.
\newblock {\em Phys. Rev. Lett.}, 117(12):121303, 2016.

\end{thebibliography}

%\appendix
%\section{An expample conner plot of parameters}
%\label{sec:app_a}
%\begin{figure}[hbt]
%  \centering
%  \includegraphics[width=1.0\textwidth]{figure/Corner.pdf}
%  \caption{Conner plot of all scanned parameters, using Runs1 10/11 as an example. The error bar of each parameter is provided.}
%  \label{fig:conner plot}
%\end{figure}

\end{document}